\documentclass[10pt,preprint]{aastex}
\usepackage[]{graphicx}
\usepackage{afterpage}

\begin{document}

\title{Direct Imaging in the Habitable Zone and the Problem of Orbital Motion}
\author{Jared R. Males, Andrew J. Skemer, Laird M. Close}
\affil{Steward Observatory, University of Arizona, Tucson, AZ 85721}
\email{jrmales@as.arizona.edu}

\begin{abstract}
High contrast imaging searches for exoplanets have been conducted on 2.4-10 m telescopes, typically at H band (1.6$\mu$m) and used exposure times of $\sim1$hr to search for planets with semi-major axes of $\gtrsim10$ AU.  We are beginning to plan for surveys using extreme-AO systems on the next generation of 30-meter class telescopes, where we hope to begin probing the habitable zones (HZs) of nearby stars. Here we highlight a heretofore ignorable problem in direct imaging: planets orbit their stars.  Under the parameters of current surveys, orbital motion is negligible over the duration of a typical observation.  However, this motion is not negligible when using large diameter telescopes to observe at relatively close stellar distances (1-10pc), over the long exposure times (10-20 hrs) necessary for direct detection of older planets in the HZ. We show that this motion will limit our achievable signal-to-noise ratio and degrade observational completeness.  Even on current 8m class telescopes, orbital motion will need to be accounted for in an attempt to detect HZ planets around the nearest sun-like stars $\alpha$ Cen A\&B, a binary system now known to harbor at least one planet.  Here we derive some basic tools for analyzing this problem, and ultimately show that the prospects are good for de-orbiting a series of shorter exposures to correct for orbital motion.  

\end{abstract}

\section{Introduction}
Orbital motion has been used in one fashion or another to detect planets around stars other than our Sun in large numbers.  The radial velocity (RV) technique monitors the Doppler shift of a stellar spectrum as the star itself orbits the planet-star center of mass, thus allowing us to infer the presence of a planet.  Similarly, the astrometry technique monitors the motion of the star on the sky and likewise infers the presence of a planet.  The transit technique monitors the reduction in brightness of the star as the orbiting planet temporarily crosses the line of sight between the telescope and the star.  

Unlike these indirect techniques, direct imaging detects light from the planet itself and spatially resolves it from the light of the star \citep{2011exop.book..111T}. The extreme difference in brightness between star and planet at small projected separations has generally limited direct imaging efforts to wide separations where orbital motion is ignorable.  The next generation of large telescopes will move us into a new regime of direct imaging, moving closer to the star.  We will even be able to begin probing the liquid water habitable zone (HZ).  Here we point out that at these tight separations orbital motion will no longer be negligible in direct imaging.  As we will show the motion of planets in the HZ (and closer), during the required integration times, will be large enough to limit our sensitivity unless we take action to correct it.

In Section \ref{sec:motivation} we present our motivation for this study and briefly review some of the related prior work.  In Section \ref{sec:quant} we develop the basic tools needed to analyze this problem, including the expected speed of orbital motion in the focal plane and the effect it has on signal-to-noise ratio ($SNR$).  In Section \ref{sec:blind} we analyze the impact orbital motion will have on a search of $\alpha$ Cen A by the Giant Magellan Telescope (GMT) working at $10\mu$m, and propose a method to mitigate this impact by de-orbiting a sequence of observations.  Then in Section \ref{sec:cued} we treat the more favorable case of a cued search, where we have prior information from an RV detection.  To do so we analyze the case of the potentially habitable planet Gl 581d being observed by the planned European Extremely Large Telescope (E-ELT).  Finally, in Section \ref{sec:conclusions}, we present our conclusions and prospects for future work.

\section{Motivation and Related Work}
\label{sec:motivation}
Moving the hunt for exoplanets into the HZ of nearby stars marks a departure from prior efforts.  Here we briefly discuss the definition of the HZ, review direct imaging results to date, discuss the differences between them and and future efforts, and finally review some closely related prior work.

\subsection{Nearby Habitable Zones}
The HZ is generally agreed to be the region around a star where a planet can have liquid water on its surface.  This is far from simply related to the blackbody equilibrium temperature, as it depends on atmospheric composition and the action of the greenhouse effect \citep{1993Icar..101..108K,2013ApJ...765..131K}, among other factors.  For our purposes it is enough to assume that the HZ is generally located at about one AU from a star, scaled by the star's luminosity
\begin{equation}
a_{HZ} \approx \sqrt{L*/L_{\sun}}\mbox{ AU}.
\end{equation}

\citet{2012ApJ...745...20T} provided three widths for the HZ based on various considerations, and then used the first 136 days of data from the \emph{Kepler} mission to estimate that the fraction of sun-like stars (spectral types FGK) with an earth-like planet in the HZ is $\eta_{\earth} \approx 0.34$.  More generally, this analysis indicates that $\eta_{planet} \approx 1.2$, implying that \emph{every sun-like star is likely to have a planet in its HZ, and some will have more than one}.  While this exciting result is based on a very large extrapolation from the earliest \emph{Kepler} results, it is currently one of our best estimates of planet frequency in the HZ.

This topic was recently brought to the fore with the announcement of $\alpha$ Cen Bb by \citet{2012Natur.491..207D}.  Discovered using the RV technique, $\alpha$ Cen Bb is an $m\sin{i} = 1.13M_{\earth}$ planet orbiting a K1 star at 0.04 AU.  While certainly not in the HZ, this discovery has exciting implications for the presence of planets in the HZ of the nearest two sun-like stars.  

The above arguments hint that planets will be common in the HZ of sun-like stars.  We are about to enter a new era of exoplanet direct imaging.  With the next generation of giant telescopes and high-performance spaced-based coronagraphs we will be searching for planets in this scientifically important region around nearby stars.  

\subsection{A Different Regime}
The typical search for exoplanets with direct imaging has used 2.4m (Hubble Space Telescope, HST) to 10m (Keck) telescopes.  These surveys have mostly concentrated on young giant planets, which are expected to be self-luminous as they dissipate heat from their formation.  This allows them to be detected at wider separations from their host stars, where reflected starlight would be too faint.  This has also caused planet searches to typically work at H band ($\sim 1.6\mu$m), with exposure times of $\sim1$ hr.  Examples conforming to these stereotypes include \citet{2005AJ....130.1845L} using HST/NICMOS; the Gemini Deep Planet Search \citep{2007ApJ...670.1367L}; the Simultaneous Differential Imaging survey using the Very Large Telescope  and MMT \citep{2007ApJS..173..143B}; the Lyot Project at the Advanced Electro-Optical System telescope \citep{2010ApJ...716.1551L}; the International Deep Planet Survey \citep{2012A&A...544A...9V}; and the Near Infrared Coronagraphic Imager at Gemini South \citep{2010SPIE.7736E..53L}.

These searches have had some success.  Examples include the 4 planets orbiting the A5V star HR 8799 \citep{2008Sci...322.1348M, 2010Natur.468.1080M}, with projected separations of 68, 38, 24, and $\sim15$ AU.  These correspond to orbital periods of $\sim460$, $\sim190$, $\sim100$, and $\sim50$ years, respectively.  The A5V star $\beta$ Pic also has a planet \citep{2010Sci...329...57L} orbiting at $\sim8.5$ AU with a period of $\sim 20$ years \citep{2012A&A...542A..41C}.  Another A star, Fomalhaut, has a candidate planet on an 872 year (115 AU) orbit \citep{2008Sci...322.1345K}.  At these wide separations it takes months, or even years, to notice orbital motion.  

In the much closer HZ, however, orbital periods will be on the order of one year.  We show in some detail that this is fast enough to yield projected motions of significant fractions of the point spread function (PSF) full width at half maximum (FWHM) over the course of an integration.  The resulting smeared out image of the planet will have a lower $SNR$, making our observations less sensitive. 

\subsection{Long Integration Times}
In addition to HZ planets having higher orbital speeds than the current generation of imaged exoplanets, integration times required to detect them will be much longer.  Direct imaging surveys to date have mostly worked in the infrared while attempting to detect young planets still cooling after formation.  The coming campaigns to image planets in the HZ of nearby stars will focus on older planets, which will be less luminous in the near infrared.  In the HZ, starlight reflected from the planet will be more important.  The result is integration times required to detect such planets will be tens of hours, rather than the $\sim 1$ hour characteristic of current campaigns.

Consider the Exoplanet Imaging Camera and Spectrograph (EPICS), an instrument proposed for the E-ELT.  \citet{2010SPIE.7735E..81K} predicted that EPICS will be able to image the RV detected planet Gl 581d, which has a semi-major axis of 0.22 AU with a period of $\sim67$ days \citep{2011arXiv1109.2505F,2012arXiv1207.4515V}.  This orbit places it on the outer edge of the HZ of its M2.5V star \citep{2011ApJ...729L..26V}. EPICS will be able to detect Gl 581d, at a planet/star contrast of $2.5\times 10^{-8}$,  in 20 hrs with $SNR = 5$ \citep{2010SPIE.7735E..81K}.  Since this is a ground based instrument, a 20 hour integration will be broken up over at least 2 nights.  Plausible observing scenarios could extend this to several nights, taking into account such things as the need for sky rotation.  As we will show, the planet will move several FWHM on the EPICS detector during a multi-day observation.     

More generally, \citet{2006A&A...447..397C} showed that when realistic non-common path wavefront errors are taken into account, the integration times required to achieve the $10^{-9}$ to $10^{-10}$ contrast necessary to detect an earth-like planet around a sun-like star approach 100 hours on the ground, even on a 100m telescope with extreme-AO and a perfect coronagraph. One of several concerns about the feasibility of a 100 hour observation from the ground is that such a long observation will be broken up over many nights.  

With net exposure times of 20 to 100 hrs, and total elapsed times for ground based observations of several to tens of days, HZ planets will move significantly over the course of a detection attempt.  The focus of this investigation is the impact of the orbital motion of a potentially detectable planet on sensitivity.

\subsection{Related Work}
Though it has not yet been a significant issue in direct imaging of exoplanets, orbital motion has been considered in several closely related contexts.  Here we briefly review a select portion of the literature. A very similar problem has been addressed in the context of searching for objects in our solar system, such as Kuiper Belt objects (KBOs), which can have proper motions on the order of 1" to 6" per hour \citep{1999AJ....118.1411C}.  Blinking images to look for moving objects by eye is a well established technique.  A more computationally intensive form of blinking images proceeds  by shifting-and-adding a series of short exposures along trial paths, usually assumed to be linear.  This ``digital tracking'' makes it possible to detect KBOs too faint to appear in a single exposure.  This has been done both from the ground \citep{1999AJ....118.1411C,2008PASJ...60..285Y} and from space with HST \citep{2004AJ....128.1364B}.  More recently \citet{2010PASP..122..549P} have taken into account nonlinear motion and optimized selection of the search space, especially important given the large data sets that facilities such as the Large Synoptic Survey Telescope will produce.  

Orbital motion is an important consideration when planning coronagraphic surveys of the HZs of nearby stars.  \citet{2005ApJ...624.1010B} treats the problem of completeness extensively. Large parts of the HZ will be within the inner working angle of the Terrestrial Planet Finder-Coronagraph (TPF-C) and so undetectable during a single observation.   Also discussed in \citet{2005ApJ...624.1010B} is photometric completeness - that is how long the TPF-C must integrate on a given star to detect an earth-like planet in the HZ.  Other work on this topic includes \citet{2010ApJ...715..122B} and \citet{2004ApJ...607.1003B}.  These analyses consider orbital motion only between observations, not during a single observation as we do here.  In general, the scenarios considered for these studies involved space-based high-performance coronagraphs on medium to large telescopes.  In such cases exposure times were short enough and continuous so that orbital motion should be negligible during a single observation.

The work most similar to our analysis here is the detection of Sirius B at $10\mu$m by \citet{2011ApJ...730...53S}, in fact, it was part of our motivation for the present study.  \citet{2011ApJ...730...53S} used the well known orbit of the white dwarf companion to Sirius to de-orbit 4 years worth of images.  Before accounting for orbital motion, Sirius B appeared as only a low $SNR$ streak, but after shifting based on its orbit it appears as a higher $SNR$ point source from which photometry can be extracted.  Similar to this method, we will analyze the prospects for de-orbiting sequences of images, only we consider the case with no prior information at all, and with orbital elements with significant uncertainties.

\section{Quantifying The Problem}
\label{sec:quant}
In this section we will quantify the effects of orbital motion on an attempt to detect an exoplanet.  Our first step will be to determine how fast planets move when projected on the focal plane of a telescope.  Then we'll illustrate the impact this motion will have on the $SNR$ and the statistical sensitivity of an observation.

\subsection{Basic Equations}
We begin by considering a focal plane detector working at a wavelength $\lambda$ in $\mu$m.  The FWHM of the PSF for a telescope of diameter $D$ in m, neglecting the central obscuration,  is
\begin{equation}
\mbox{FWHM} = 0.2063\frac{\lambda}{D}\mbox{ arcsec}.
\end{equation}
If we are observing a planet in a face-on circular (FOC) orbit with a semi-major axis of $a$ in AU at distance $d$ in pc, its angular separation will be $a/d$ arcsec.  At the focal plane the projected separation will then be
\begin{equation}
\rho = 4.847\frac{a D}{\lambda d} \mbox{ in FWHM}.
\end{equation}
We note that it will occasionally be convenient to specify $\rho$ in AU instead of FWHM.  When it is not clear from the context we will use the notation $\rho_{au}$ to denote this.  

The orbital period is $P=365.25\sqrt{a^3/M_*}$ days around a star of mass $M_*$ in $M_{\sun}$.  In one period, the planet will move a distance equal to the circumference of its orbit, $2\pi\rho$, so the speed of the motion in a FOC orbit will be\footnote{This result is equivalent to defining the gravitational constant in the focal plane as $G=(0.0834D/(\lambda d))^2$ and using the equation for speed in a circular orbit $v_{circ} = \sqrt{\frac{GM_*}{a}}$.}
\begin{equation}
v_{FOC} = 0.0834\left(\frac{D}{1\mbox{m}}\right) \left(\frac{1\mu\mbox{m}}{\lambda}\right) \left(\frac{1\mbox{pc}}{d}\right) \sqrt{ \left(\frac{M_*}{1M_\sun}\right)\left(\frac{1\mbox{AU}}{a}\right)} \mbox{ in FWHM day}^{-1}.
\label{eqn:vfoc}
\end{equation}

In the general case, the equations of motion in the focal plane are
\begin{eqnarray}
\dot{x}&=&v_{FOC}\sqrt{\frac{1}{1-e^2}}\Big[e\sin(f)\left(\cos(\Omega)\cos(\omega+f)-\sin(\Omega)\sin(\omega+f)\cos(i)\right) \nonumber \\
& &-(1+e\cos(f))\left(\cos(\Omega)\sin(\omega+f)+\sin(\Omega)\cos(\omega+f)\cos(i)\right)\Big] \nonumber\\
\dot{y}&=&v_{FOC}\sqrt{\frac{1}{1-e^2}}\Big[e\sin{(f)} \left(\sin{(\Omega)}\cos(\omega+f)+\cos(\Omega)\sin(\omega+f)\cos(i)\right)\\
& &-(1+e\cos(f))\left(\sin(\Omega)\sin(\omega+f)-\cos(\Omega)\cos(\omega+f)\cos(i)\right)\Big] \nonumber \\
v_{om} &=& \sqrt{\dot{x}^2+\dot{y}^2} \nonumber
\end{eqnarray}
where $\Omega$ is the longitude of the ascending node, $\omega$ is the argument of pericenter, $i$ is the inclination, and the true anomaly $f$ depends on $a$, $e$, and the time of pericenter passage $\tau$ through Kepler's equation \citep{2010exop.book...15M}.

\begin{figure}
\begin{center}
\includegraphics[scale=.64]{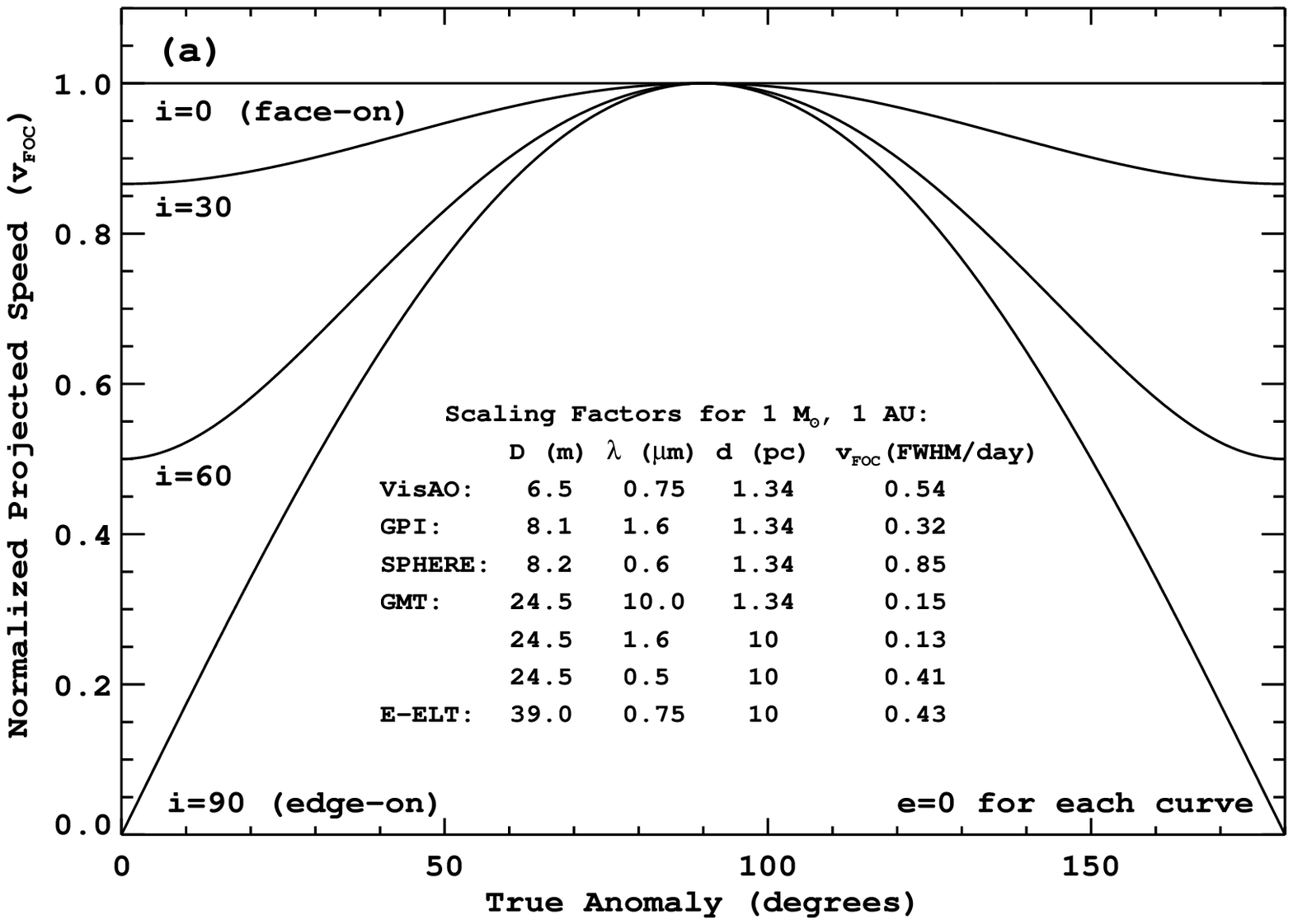}
\includegraphics[scale=.64]{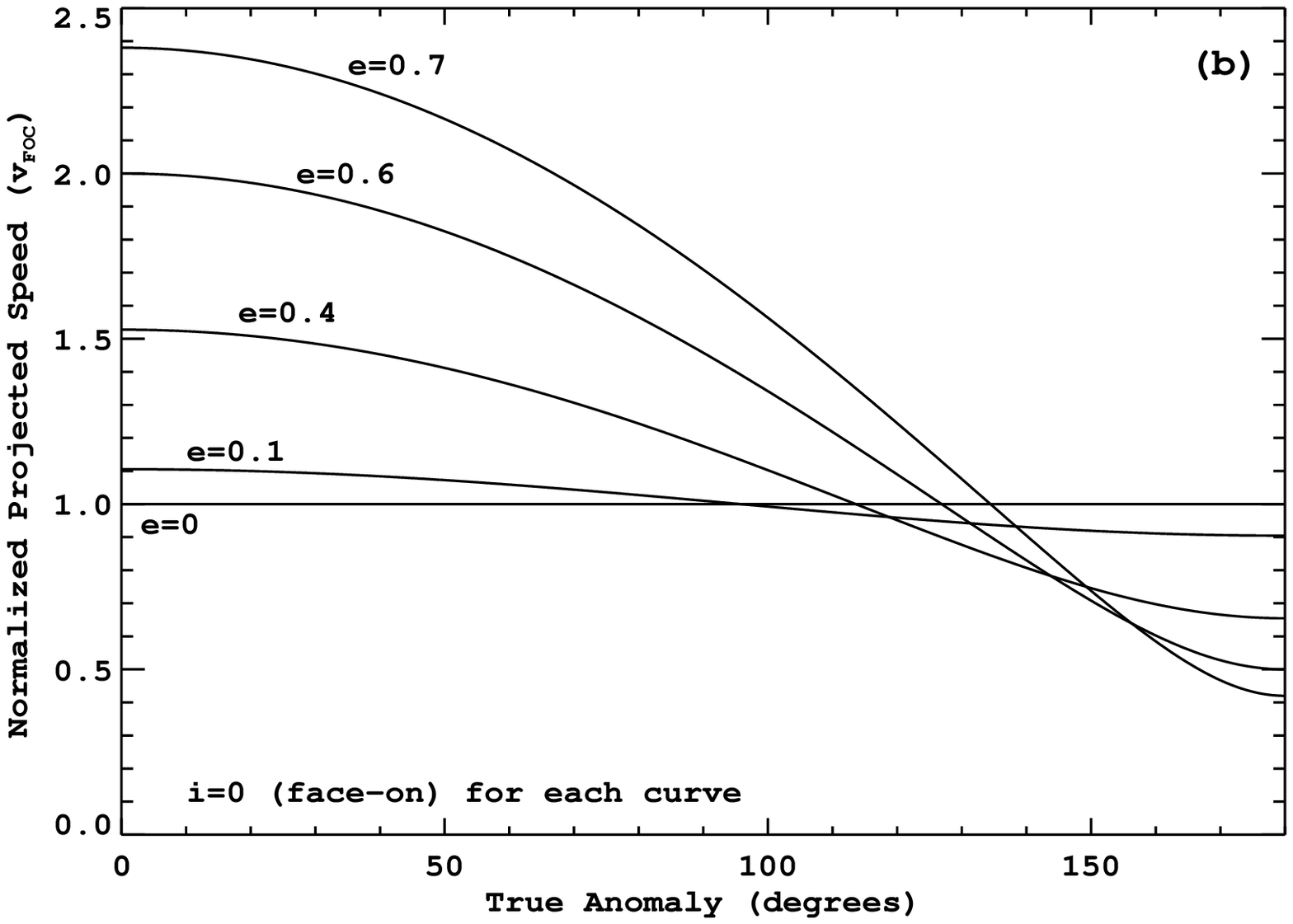}
\end{center}
\caption{
Magnitude of projected orbital speed, normalized to 1 FWHM day$^{-1}$, for 1 AU orbits around a $1M_\sun$ star.  In (a) we show the orbital speeds for circular orbits at various inclinations, and in (b) we show the speeds for face-on orbits at various eccentricities.  We give scaling factors in (a) for MagAO/VisAO \citep{2012SPIE.8447E..0XC}, GPI \citep{2012SPIE.8446E..1UM}, SPHERE/ZIMPOL \citep{2010SPIE.7735E.144R}, GMT \citep{2012SPIE.8444E..1HJ}, and E-ELT/EPICS \citep{2010SPIE.7735E..81K}.    These scalings can be applied to the y-axis of either plot for various scenarios.  These cases can also be scaled for different semi-major axes, telescopes, wavelengths, star masses and distances, by $v_{FOC} \propto \frac{D}{\lambda d}\sqrt{\frac{M_*}{a}}$. See the text for the general equations of motion for arbitrarily oriented eccentric orbits. 
\label{fig:norm_motion}}
\end{figure}

In Figure \ref{fig:norm_motion} we show the variation in projected orbital speed for both circular orbits at several inclinations, and face-on eccentric orbits ($i=0$), for a planet orbiting a $1M_\sun$ star at 1 AU. In the plots we normalized speed to 1, and provide $v_{FOC}$ for several interesting cases.  These various scenarios produce projected orbital speeds of appreciable fractions of a FWHM per day.  We will later show that, especially for ground based imaging, this causes a significant degradation in our sensitivity.

\afterpage{\clearpage}

Our main focus here is on planets in the HZ.  Our simple definition of the HZ results in $a_{HZ} \propto \sqrt{L_*}$.  Now, on the main sequence mass and luminosity approximately follow scaling laws of the form $L_* \propto M_*^b$, where $b > 2$ except for very massive stars. So according to Equation (\ref{eqn:vfoc}) we expect $v_{FOC}$ in the HZ to increase as $M_*$ decreases, i.e. M stars will have faster HZ planets than G stars.  For example, a planet in the HZ of $\alpha$ Cen B ($M_* = 0.9M_\sun$, $L_* = 0.5L_\sun$) will be moving roughly 20\% faster than a planet in the HZ of $\alpha$ Cen A ($M_* = 1.1M_\sun$, $L_* = 1.5L_\sun$) (stellar parameters from \citet{2010MNRAS.405.1907B}).

To provide a more concrete example we return to the 20 hour observation of Gl 581d by the E-ELT/EPICS proposed by \citet{2010SPIE.7735E..81K}.  Using a wavelength of $0.75\mu m$ with Equation (\ref{eqn:vfoc}) we find $v_{FOC} = 0.82$ FWHM per day, or a total of 0.68 FWHM for a continuous 20 hour observation.  Since this is a ground based observation the actual amount of motion to consider is $\sim1.15$ FWHM over the $\sim1.4$ days minimum it would take to integrate for 20 hours.  Were this a face-on orbit, an eccentricity of 0.25 \citep{2011arXiv1109.2505F} would increase the maximum orbital speed to as much as 1.05 FWHM per day, or 1.47 FWHM minimum for a 20 hour ground based observation.  

\subsection{Impact on Signal-to-Noise Ratio}
So what does the orbital motion calculated above do to our observations?  To find out we consider a simple model of aperture photometry.  Let us assume that we are conducting aperture photometry with a fixed radius $r_{ap}$, that the PSF is Gaussian, and that we are limited by Poisson noise from a photon flux $N$ per unit area.  With these assumptions, the optimum $r_{ap}$ is 0.7 FWHM, but taking into account centroiding uncertainty $r_{ap}\approx 1$ FWHM is typical.  We will approximate orbital motion at speed $v_{om}$ by substituting $x \rightarrow x - v_{om}t - x_0$.  Orbits are of course not linear, but this will be approximately valid over short periods of time.  The parameter $x_0$ allows us to optimize the placement of the aperture to obtain the maximum signal, i.e. centering the aperture in the planet's smeared out flux.  Note that with the exception of this centering parameter, this model appears quite naive in that we are not adapting the aperture radius and are pretending that we won't notice a smeared out streak in our images.

Now the $SNR$ in the fixed-size aperture after time $\Delta t$ will be
\begin{equation}
SNR_{fix} = \frac{\displaystyle \int_{0}^{r_{ap}} \int_{0}^{2\pi} \int_0^{\Delta t} I_0 e^{\left(-4 \ln 2 ((r\cos\theta - v_{om}t - x_0)^2 + r^2\sin^2\theta)\right)} dt d\theta rdr.}
{\sqrt{N\pi r_{ap}^2 \Delta t_{int}}}
\label{eqn:s_g}
\end{equation}
where $I_0$ is the peak value of the PSF.  In the case of no orbital motion $v_{om} = 0$ and aperture $r_{ap} = 1$ FWHM, so we have
\begin{equation}
SNR_o = \frac{0.6 I_0 \sqrt{\Delta t}}{\sqrt{N}}.
\end{equation}
As a simple alternative to a fixed size aperture, we also consider allowing our photometric aperture to expand along with the motion of the planet.   This aperture will collect the same signal as in $SNR_o$, but the noise increases with the area as $2r_{ap}v_{om}\Delta t$, so we have
\begin{equation}
SNR_{exp} = \frac{0.6 I_0 \sqrt{\Delta t}}{\sqrt{N\left(1+(2/\pi)v_{om}\Delta t\right)}}.
\end{equation}
A convenient scaling is to multiply top and bottom by $\sqrt{v_{om}}$ and work in normalized $SNR$ units of $I_o/\sqrt{Nv_{om}}$.  This puts time in terms of FWHM of motion, $\epsilon = v_{om}\Delta t$, and allows comparisons without specifying $v_{om}$.  

\begin{figure}
\begin{center}
\includegraphics[scale=.64]{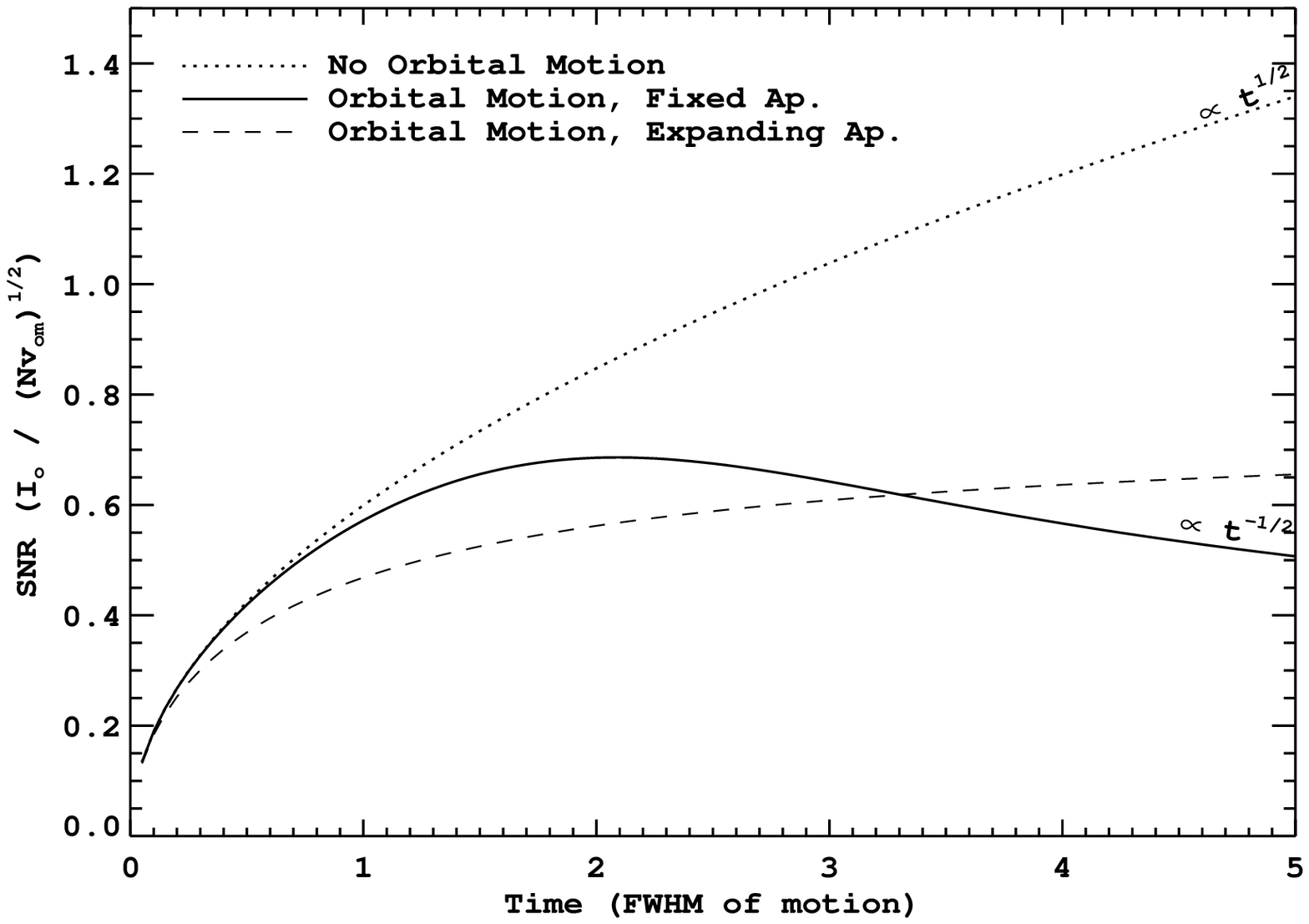}
\includegraphics[scale=.64]{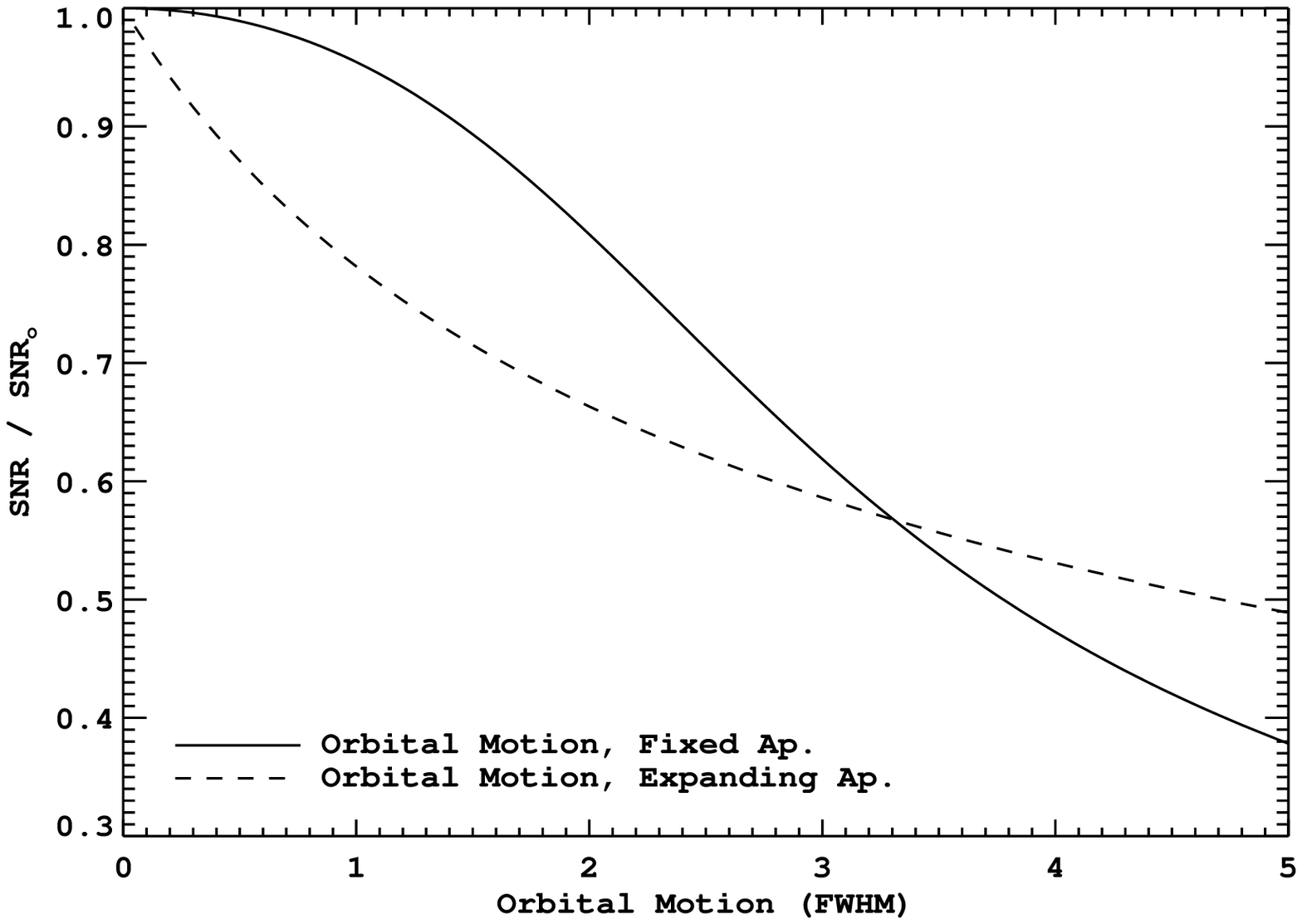}
\end{center}
\caption{Top panel: $SNR$ of a Gaussian PSF with and without orbital motion, in normalized units with time given as FWHM of motion.  With no orbital motion $SNR_o \propto \sqrt t$.  Equation (\ref{eqn:s_g}) was used to calculate the $SNR$ with orbital motion.  After $\sim 2$ FWHM of movement, a maximum is reached and the observation can only be degraded by integrating further.  Note that the fixed-aperture orbital motion case eventually goes down as $SNR \propto 1/\sqrt t$.  For comparison we also show the results with an aperture expanding with the moving planet, which eventually reaches a limit of 0.75.  In the bottom panel we show the fractional reduction in $SNR$ due to orbital motion for the fixed radius photometric aperture.  
\label{fig:snr_orbmotion}}
\end{figure}

\afterpage{\clearpage}

In Figure \ref{fig:snr_orbmotion} we plot the normalized $SNR$ vs. time (measured in terms of FWHM of motion) with and without orbital motion and for both the fixed and expanding aperture cases. For the fixed aperture, after $\sim 2$ FWHM of orbital motion a maximum of 0.69 is reached, and from there noise is added faster than signal.  This means that further integration only degrades the observation.  

The expanding aperture $SNR_{exp}$ exceeds the maximum of $SNR_{fix}$ after about 8 FWHM of motion, and

\begin{equation}\displaystyle\lim_{x \to \infty} 0.6 \frac{\sqrt{x}}{\sqrt{1+(2/\pi)x}} = 0.6\sqrt{\frac{\pi}{2}} \approx 0.75.\end{equation}

So if we integrate 4 times longer, adjusting the aperture size would allow us to gather a little more $SNR$, but only to a point.  Given this large increase in telescope time for a relatively small improvement in $SNR$ (only $\sim9\%$ even if we integrate forever), and its better performance for smaller amounts of motion, the fixed-radius aperture will be our baseline for further analysis -- keeping in mind that in some cases it may not be the true optimum.  

The peak in $SNR_{fix}$ (equation \ref{eqn:s_g}) sets the maximum nominal integration time before orbital motion will prevent us from achieving the science goal.   That is $\Delta t_{max} = (SNR_{max}/0.6)^2.$  If the observation of a \emph{stationary planet} would require an integration time longer than $\Delta t_{max}$, then we can't achieve the desired $SNR$ on an \emph{orbiting planet}.  This also sets the maximum orbital motion $\epsilon_{max} = v_{om}\Delta t_{max}.$  From Figure \ref{fig:snr_orbmotion} we find that $\epsilon_{max} = 1.3$ FWHM.  If more than 1.3 FWHM of motion occurs during an observation, we will not achieve the required $SNR$.

We also show the fractional reduction in $SNR$ in Figure \ref{fig:snr_orbmotion}.  Almost no degradation occurs until after $\sim 0.2$ FWHM of motion has occurred.  $SNR$ is reduced by $\sim 1\%$ after 0.5 FWHM of motion, $\sim 5\%$ after 1.0 FWHM, and by $\sim 19\%$ after 2.0 FWHM of motion. We must now decide how much $SNR$ loss we can accept in our observation.  

\begin{figure}
\begin{center}
\includegraphics[scale=.64]{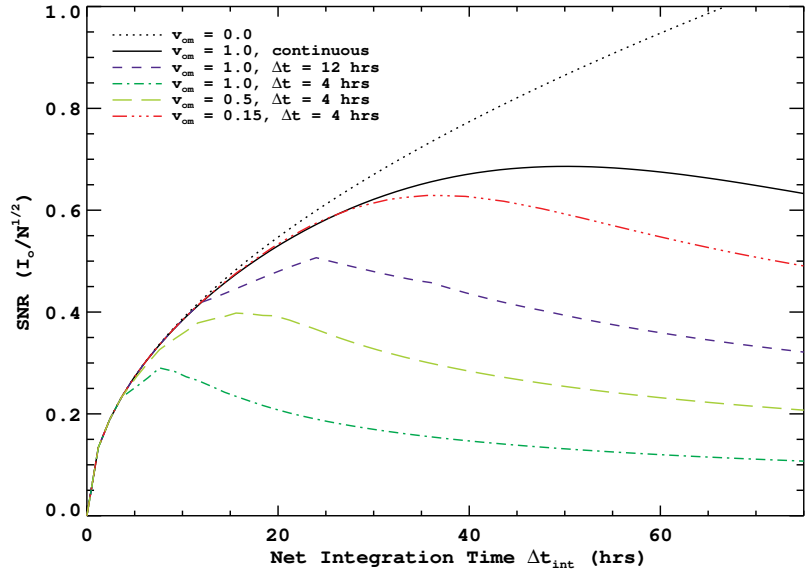}
\end{center}
\caption{Here we show the impact of orbital motion when combined with finite nightly integration times.  The $SNR$ of a Gaussian PSF with and without orbital motion is plotted in arbitrary units vs $\Delta t_{int}$.  The orbital speed $v_{om}$ is given in FWHM day${^{-1}}$.
\label{fig:snr_orbmotion_gndbase}}
\end{figure}


\afterpage{\clearpage}

The above analysis assumes a continuous integration.  On a ground-based telescope one must consider that the maximum continuous integration time is $\lesssim 12$ hours, and in practice will likely be much shorter when performing high contrast AO corrected imaging.  For instance, an exposure of 20 hours might have to be broken up over 4 or 5 or more nights, when considering the vagaries of seeing (required AO performance), airmass (either through transmission or $r_0$ requirements), rotation rate (for ADI), and weather.  We can adapt the calculations for a ground based integration as follows
\begin{equation}
SNR_{gnd} = \frac{\displaystyle\int_{0}^{r_{ap}} \int_{0}^{2\pi} \left[ \displaystyle\sum_{j=1}^{j=M} \int_{t_j}^{t_j + \Delta t_j} I_0 e^{\left(-4 \ln 2 ((r\cos\theta - v_{om}t - x_0)^2 + r^2\sin^2\theta)\right)} dt \right] d\theta rdr.}
{\sqrt{N \pi r_{ap}^2 \Delta t_{int}}}
\label{eqn:s_gnd}
\end{equation}
In this expression we have broken the observation up into M integration sets which start at times $t_j$ and have lengths $\Delta t_j$. The total integration time is $\Delta t_{int} = \displaystyle\sum_{j=1}^{j=M} \Delta t_j$ and the total elapsed time of the observation is $\Delta t_{tot} = t_M+\Delta t_M - t_1$.

We plot the results for a few ground-based scenarios in Figure \ref{fig:snr_orbmotion_gndbase}.  As one can see, observations of planets with orbital motion will be significantly degraded from the ground.  This problem, which has been negligible in the high contrast planet searches to date, only becomes worse as we consider larger telescopes and improvements in AO technology which allow searches at shorter wavelengths.  We next analyze how this reduction in $SNR$ will affect our ability to detect exoplanets by increasing the rate at which spurious detections occur.

\subsection{Impact on Statistical Sensitivity}

Now we turn to the problem of detecting a planet of a given brightness.  A planet is considered detected if its flux is above some threshold $SNR_t$, which is chosen for statistical significance.  The goal in choosing this threshold is to detect faint planets while minimizing the number of false alarms. For the purposes of this analysis we assume Gaussian statistics, in which case the false alarm probability ($P_{FA}$) per trial is
\begin{equation}
P_{FA} = \frac{1}{2}\mbox{erfc}\left(\frac{SNR}{\sqrt{2}}\right)
\label{eqn:gauss_pfa}
\end{equation}
Typically, planet hunters use a threshold of $SNR = 5$, which gives $P_{FA}=2.9\times10^{-7}$.  The number of false alarms per star, the false alarm rate ($FAR$), is then
\begin{equation}
FAR = P_{FA}\times N_{trials}.
\end{equation}
where $N_{trials}$ is the number of statistical trials per star. Following \citet{2008ApJ...673..647M}, for a stationary planet $N_{trials}$  is just the number of photometric apertures in the image. A typical Nyquist sampled detector of size 1024x1024 pixels has $N_{trials} \sim 8\times10^4$.  Thus, an $SNR=5$ threshold will result in $FAR \sim 0.02$ -- about 1 false alarm for every 50 observations.  In the speckle limited case with non-Gaussian statistics, $FAR$ will be worse than this for the same $SNR$ \citep{2008ApJ...673..647M}.  In any case, the $FAR$ is the statistic which determines the efficiency of a search for exoplanets with direct imaging.  A high $FAR$ will cause us to waste telescope time following up spurious detections, while raising the $SNR$ threshold to counter this limits the number of real planets we will detect. 

The reduction of $SNR$ caused by orbital motion confronts us with three options.  Option I is to maintain the detection threshold constant and accept the loss of sensitivity. Option II is to lower the detection threshold to maintain sensitivity, accepting the increase in FAR.  Option III is to correct for orbital motion, which as we will show also causes an increase in FAR.

\begin{figure}
\begin{center}
\includegraphics[scale=.64]{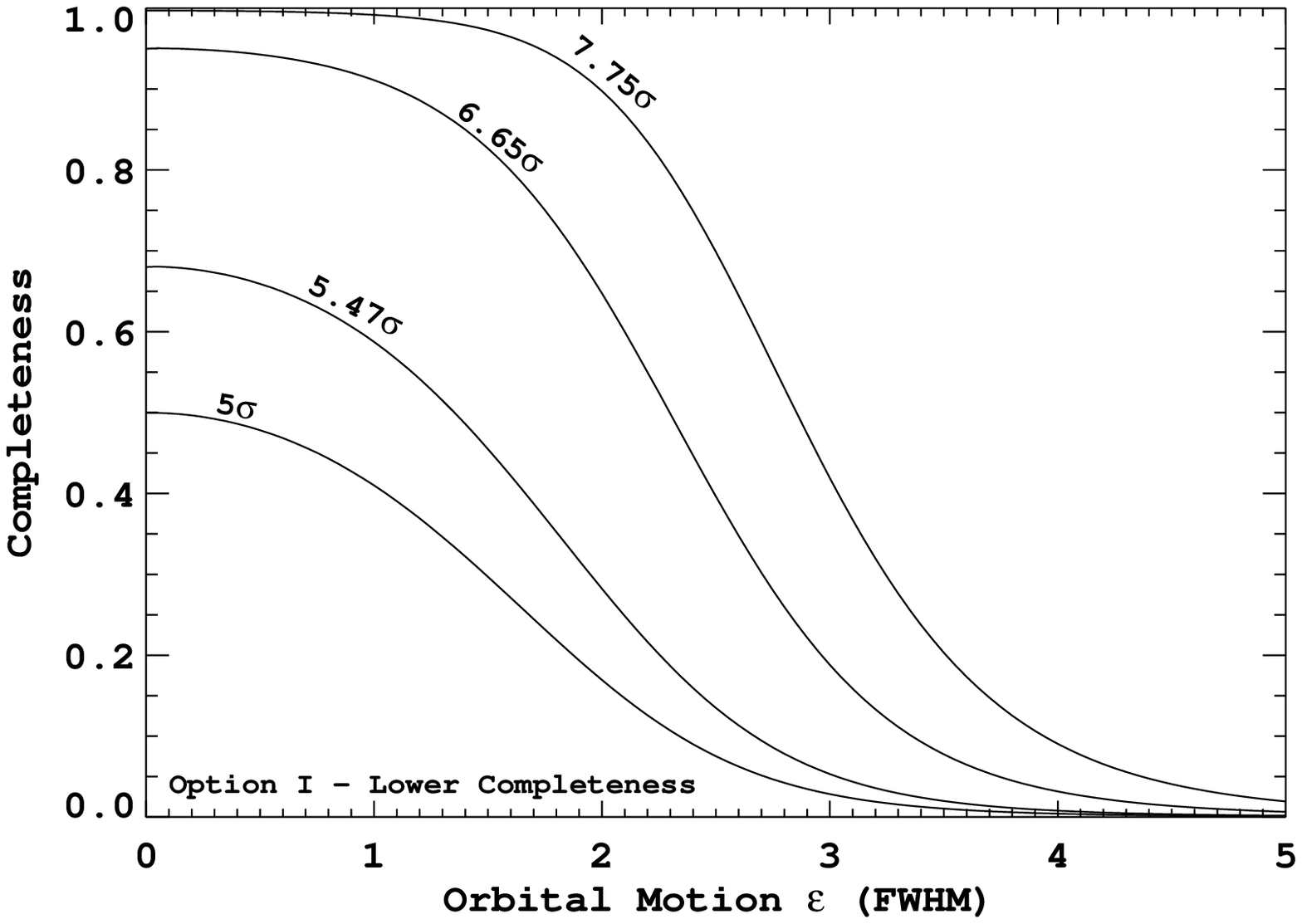}
\includegraphics[scale=.64]{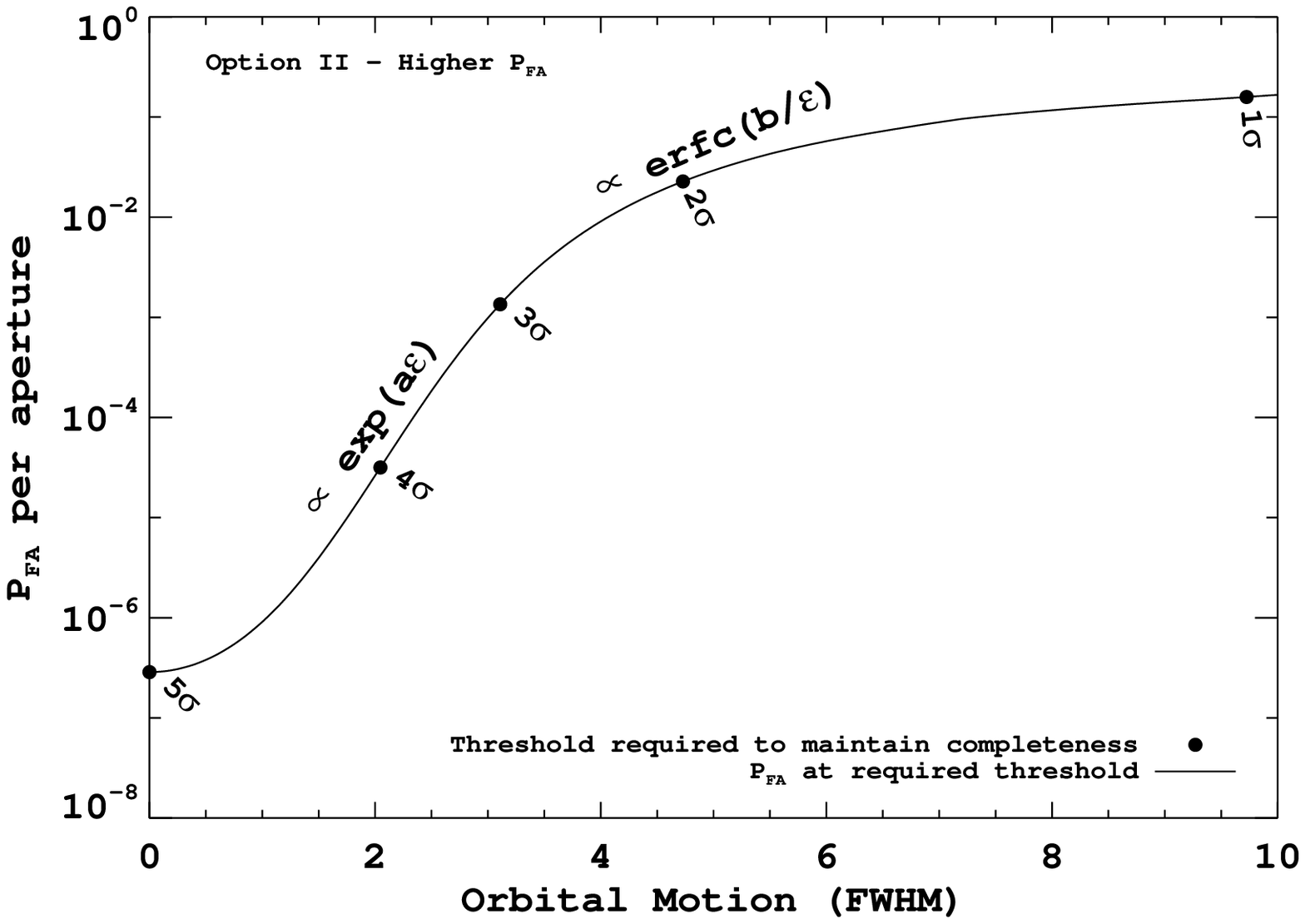}
\end{center}
\caption{Top panel: completeness as a function of orbital motion if we maintain our detection threshold at $5\sigma$.  Planet brightness is expressed as the $SNR$ at which we would be 50\%($5\sigma$), 68\%($5.47\sigma$), 95\%($6.65\sigma$), and 99.7\% ($7.75\sigma$) complete with no orbital motion.  Bottom panel: the increase in false alarm probability $(P_{FA})$  if we lower the detection threshold to maintain 50\% completeness for an orbiting planet that would have a brightness of $5\sigma$ were it stationary.  After $\sim 1$ FWHM of motion $P_{FA}$ increases exponentially until $\sim 4$ FWHM where it becomes asymptotic to 0.5.
\label{fig:option_I_II}}
\end{figure}

\afterpage{\clearpage}

\subsubsection{Option I: Do Nothing}
The default option is to do nothing, keeping our detection threshold set as if orbital motion is not significant.  The drawback to this is that we will detect fewer planets.  To quantify this we use the concept of completeness, that is the fraction of planets of a given brightness we detect.  For Gaussian statistics and detection threshold $SNR_t = 5$, the search completeness is given by 
\begin{equation}
C(\epsilon) = 1-\frac{1}{2}\mbox{erfc}\left(\frac{SNR(\epsilon) - 5}{\sqrt{2}}\right).
\end{equation}
where $\epsilon = v_{om}\Delta t$ is the amount of motion.  In Figure \ref{fig:option_I_II} (top) we show the impact of orbital motion on search completeness.  Maintaining the detection threshold lowers completeness.  How much depends on the completeness level, with brighter planets being less affected.  For planets bright enough to yield 95\% completeness with no motion, significant reduction in the number of detections begins after $\sim 1$ FWHM of motion.  For 99.7\% completeness the impact becomes significant after $\sim 1.5$ FWHM.  

\subsubsection{Option II: Lower Threshold}

Once orbital motion is recognized to be significant, a simple countermeasure would be to lower the detection $SNR$ threshold in order to maintain completeness.  The drawback to this option is that we have more false alarms, which must then be followed up using more telescope time.  This results in a less efficient search.  In Figure \ref{fig:option_I_II} (bottom) we show $P_{FA}$ as a function of orbital motion, and denote the detection threshold we must use to maintain 50\% completeness for a planet bright enough to give $SNR = 5$ were it stationary.  Note that $P_{FA}$ begins to increase exponentially after $\sim 1$ FWHM of motion.  After $\sim 4$ FWHM $P_{FA}$ begins approaching 0.5 asymptotically.  Once $\epsilon \approx 2$ FWHM the number of false alarms per 1024x1024 image approaches 1.

\subsubsection{Option III: De-orbit}
\label{sect:deorbit}

Option III is to correct for orbital motion, hoping to maintain sensitivity while limiting the increase in $P_{FA}$.  The essence of any such technique will be calculating the position of the planet during the observation, and de-orbiting in some way, say shift-and-add (SAA) on a sequence of images.  The drawback of this approach is that it will produce more false alarms per observed star due to the increased number of trials, similar to lowering the detection threshold.  If the orbit were precisely known, we could proceed with almost no impact on $FAR$.  However, in the presence of uncertainties in orbital parameters or in a completely blind search we will have to consider many trial orbits.  For now we can perform a ``back-of-the-envelope'' estimate of the number of possible orbits to understand how much $FAR$ will increase.  To do so, we begin by placing bounds on the problem.

We can first establish where on the detector we must consider orbital motion.  At any separation $r$ from the star, the slowest un-bound orbit will have the escape velocity.  Since we know that physical separation is greater than or equal to projected separation, $r \ge \rho$, and that maximum projected speed will occur for inclination $i=0$, we know that 
\begin{equation}
v_{esc} = \sqrt{2}v_{FOC}(a \rightarrow \rho)
\end{equation}
sets the upper limit on the projected focal plane speed of an object in a bound orbit.  We can also set an upper limit on the amount of motion $\epsilon_{max}$ we can tolerate over the duration $\Delta t_{tot}$ of the observation based on the $SNR$ degradation it would cause.  So we only need consider orbital motion when
\begin{equation}
\sqrt{2}v_{FOC}(\rho)\Delta t_{tot} > \epsilon_{max}.
\end{equation}
From here we determine the upper limit on projected separation from the star for considering this problem:
\begin{equation}
\rho_{max} = 0.0136M_* \left(\frac{D}{\lambda d} \frac{\Delta t_{tot}}{\epsilon_{max}}\right)^2 \mbox{ AU.}
\label{eqn:rhomax}
\end{equation}
By the same logic, for any point closer than $\rho_{max}$ the maximum possible change in position is
\begin{equation}
\Delta\rho_{max} \approx \sqrt{2}v_{FOC}(\rho)\Delta t_{tot} \mbox{ in FWHM}.
\label{eqn:deltarhomax}
\end{equation}
Then we must evaluate possible orbits ending anywhere in an area of $\pi (\Delta\rho_{max})^2$ FWHM$^2$ around an initial position.

These two limits set the statistical sensitivity of an attempt to de-orbit an observation.  The number of different orbits, $N_{orb}$, will be determined by the area of the detector where orbital motion is non-negligible, and the size of the region around each point that we consider.  That is
\begin{equation}
N_{orb} \propto \int_{0}^{\rho_{max}} \Delta\rho_{max}^2 \rho d\rho.
\end{equation}
so
\begin{equation}
N_{orb} \propto \left(\frac{M_*}{\epsilon}\right)^2\left(\frac{D}{\lambda d}\right)^4 \Delta t_{tot}^4.
\end{equation}

In general $N_{trials} \propto N_{orb}$, so $FAR \propto P_{FA} \times N_{orb}$.  Larger $D$, shorter $\lambda$, closer $d$, and smaller acceptable orbital motion $\epsilon$ will then all increase $FAR$\footnote{Assuming background limited photometry with a diffraction limited PSF, we expect $\Delta t \propto 1/D^4$ \citep{1998aoat.book.....H}.  All else being equal, larger telescopes are better when considering this problem}.  Perhaps the most important feature of this result is that $N_{orb} \propto \Delta t_{tot}^4$ -- {\bf increasing integration time rapidly increases the $FAR$ of a blind search}.  Note that this is still less severe than the exponential increase in $P_{FA}$ found for merely lowering the threshold.  In the next section we will test these relationships after fully applying orbital mechanics, and see that they hold.  

\section{Blind Search: Recovering SNR after Orbital Motion}
\label{sec:blind}
In this section we consider in detail a blind search, i.e. an observation of a star for which we have no prior knowledge of exoplanet orbits.  We showed above that the problem is well constrained.  Here we derive several ways to further limit the number of trial orbits we must consider.  After that, we describe an algorithm for determining the orbital elements that must be considered and then discuss the results.  Finally, we use this algorithm to de-orbit a sequence of simulated images and analyze the impact of correlations between trial orbits on $FAR$.

To provide numerical illustrations throughout this section  we consider the problem of a 20 hour observation of $\alpha$ Cen A using the GMT at 10$\mu$m.  This scenario is loosely based on performance predictions made for the proposed TIGER instrument, a mid-IR diffraction limited imager for the GMT \citep{2012SPIE.8446E..1PH}.  The details of these predictions are not important for our purposes, so we will only assert that this is a plausible case.  There are other examples in the literature with similar integration times, such as the EPICS prediction we discussed earlier.

We assume that this 20 hr observation is broken up into five $\Delta t = 4$ hr exposures, spread over 7 nights or $\Delta t_{tot} = 6.2$ elapsed days from start to finish.  The choice of $\Delta t$ is essentially arbitrary, but we have good reasons to expect it to be shorter than an entire night.  An important consideration is the planned use of ADI, and the attendant need to obtain sufficient field rotation in a short enough time to provide good PSF calibration while avoiding self-subtraction \citep{2006ApJ...641..556M}.  The effect of airmass on seeing through $r_0 \propto \cos(z)^{3/5}$, where z is the zenith angle, and hence on AO system performance, could also cause us to observe as near transit as possible.  Efficiency will be affected by chopping and nodding, necessary for background subtraction at $10\mu$m. This will limit the net exposure time obtainable in one night.. 

Few ground-based astronomers would object to an assertion that we loose 2 nights out of 7 to weather.  We could be observing in queue mode, such that these observations are only attempted when seeing is at least some minimal value, or precipitable water vapor is low.  One can even imagine the opposite case at $10\mu$m, such that nights of the very best seeing are devoted to shorter wavelength programs.  While this scenario may be somewhat contrived, we feel that it is both plausible and realistic.  We now proceed to describe a technique that would mitigate the effects of orbital motion for our GMT example and should be applicable to other long exposure cases.

\subsection{Limiting Trial Orbits}
Here we derive limits on the semi-major axis and eccentricity of trial orbits to consider.  These limits are based only on the amount of orbital motion tolerable for the science case, and do not represent physical limits on possible orbits around the star.

It is always true that $r \ge \rho$.  This implies that, for any orbit, the separation of apocenter must obey $r_a \ge \rho$.  This allows us to set a lower bound on $a$, $a_{min}$, given a choice of $e$ through
\begin{equation}
\rho_{au} \leq a_{min}(1+e)
\end{equation}
which gives
\begin{equation}
a_{min} = 0.2063\frac{\lambda d \rho}{D(1+e)}.
\label{eqn:a_min}
\end{equation}

The fastest speed in a bound planet's orbit will occur at pericenter, and using the maximum tolerable motion $\epsilon_{max}$ during our observation of total elapsed time $\Delta t_{tot}$ we can set an upper bound on $a$ by noting that
\begin{equation}
v_{FOC}(a_{max})\sqrt{\frac{1+e}{1-e}}\Delta t_{tot} \leq \epsilon_{max}
\end{equation}
which leads to
\begin{equation}
a_{max} = \left(0.0834\frac{D}{\lambda d}\right)^2\frac{1+e}{1-e}M_*\left(\frac{\Delta t_{tot}}{\epsilon_{max}}\right)^2.
\label{eqn:a_max}
\end{equation}

Using the GMT example: for $e=0.0$, $a_{max} = 3.9$ AU; and for $e = 0.5$,  $a_{max} = 11.8$ AU.  Using Equation \ref{eqn:rhomax} we have a projected separation limit of $\rho_{max} = 7.7$ AU, so it is possible for these definitions to produce $a_{max} < a_{min}$ for certain choices of $e$ at a given $\rho$.  This condition tells us that at such a value of $e$ no orbits can move fast enough to warrant consideration.  Thus we can set a lower limit on $e$ at projected separation $\rho$
\begin{equation}
e_{min} = \frac{1}{2}\sqrt{\xi^2+8\xi} - 1 - \frac{\xi}{2}
\label{eqn:e_min}
\end{equation}
where we have simplified by pulling out
\begin{equation}
\xi = 29.66\frac{\rho}{M_*}\left(\frac{\epsilon}{\Delta t}\right)^2\left(\frac{\lambda d}{D}\right)^3.
\end{equation}

In practice, we might consider eccentricity ranges with $e_{max}$ less than 1, thus improving our sensitivity.  Inputs to our choice of $e_{max}$ could include some prior distribution of eccentricities, or dynamical stability considerations in binary star systems and systems with known outer companions.

\subsection{Choosing Orbital Elements}
\label{sect:algo}
Now we describe an algorithm for sampling the possible trial orbits over a set of $M$ sequential images.  For now, we assume no prior knowledge of orbital parameters.  We will employ a simple grid search through the parameter space bounded as described above.
\begin{enumerate}
\item Determine the region around the star to consider using Eq. (\ref{eqn:rhomax}).
\item Identify regions of interest.  In the best cases the orbital motion will be small enough that we will be able stack the images and search the result for regions with higher $SNR$ (e.g. $SNR$ $> 4$) and limit further analysis to those areas.  In the worst cases orbital motion will be large enough that we will need to blindly apply this algorithm at each pixel within the bounding region identified in the previous step.  In the present GMT-$\alpha$Cen example we are in the former case.
\item For each region, choose a size, perhaps based on $v_{esc}$ (as in Eq. \ref{eqn:deltarhomax}). \label{item:region}
\item Chose a starting point $(x_1, y_1)$, with $\rho_1 = \sqrt{x_1^2 + y_1^2}$. If we are proceeding pixel by pixel, then $(x_1, y_1)$ describes the current pixel. \label{item:startpoint}
\item Choose $e \in e_{min}(\rho_1) \ldots e_{max}$ using Equation (\ref{eqn:e_min}) and assumptions about $e_{max}$. \label{item:starte}
\item Choose $a \in a_{min}(\rho_1, e) \ldots a_{max}(e)$ using Equations (\ref{eqn:a_min}) and (\ref{eqn:a_max}).
\label{item:starta}
\item Choose time of pericenter $\tau \in t_1-P(M_*, a) \ldots t_1$ where $P$ is the orbital period and $t_1$ is the time of the first image. Now calculate the true anomaly $f(t_1;a,e,\tau,P)$ using \emph{Kepler}'s equation and physical separation using:
\begin{equation}
r = \frac{a(1-e)}{1+e\cos(f)}
\end{equation}
\label{item:starttau}
\item if $e \neq 0$: Choose $\omega \in 0 \ldots 2\pi$ \label{item:choose_omega}\\
if $e = 0$: set $\omega = 0$.
\label{item:calcelements}
\item if $\sin(\omega+f) > 0$:
\begin{enumerate}

\item Given $e$, $a$, $\tau$, $f$, and $\omega$, calculate 
\begin{equation}
\cos i = \frac{\pm\sqrt{\frac{\rho^2}{r^2} - \cos^2(\omega+f)}}{\sin(\omega+f)}
\end{equation}
\begin{equation}
\sin \Omega = \frac{y\cos(\omega + f) - x\sin(\omega+f)\cos i}{r(\cos^2(\omega+f)+\sin^2(\omega+f)\cos^2 i)}
\end{equation}
\begin{equation}
\cos \Omega = \frac{y\sin(\omega+f)\cos i +  x\cos(\omega + f) }{r(\cos^2(\omega+f)+\sin^2(\omega+f)\cos^2 i)}
\end{equation}
where $\Omega$ should be determined in the correct quadrant.
\label{item:iOmega}
\item We now have a complete set of elements, and so can SAA the sequence of images based on these orbits (one for each $i$).  Doing so requires calculating the true anomaly $f_j$ at the time of each image, and then calculating the projected orbital position of the prospective companion in each image.
\label{item:SAA}
\end{enumerate}
\item if $\sin(\omega + f) = 0$, we do not have a unique solution for inclination.  This is the special case where the planet is passing through the plane of the sky. 
\label{item:eneg0_sineq0}
\begin{enumerate}
\item for $\omega + f = 0$ calculate $\Omega$:
\begin{equation}
\sin \Omega = \frac{y}{r}
\end{equation}
\begin{equation}
\cos \Omega = \frac{x}{r}
\end{equation}
or for $\omega + f = \pi$ calculate $\Omega$:
\begin{equation}
\sin \Omega = \frac{-y}{r}
\end{equation}
\begin{equation}
\cos \Omega = \frac{-x}{r}
\end{equation}
determining $\Omega$ in the correct quadrant.
\item Choose $i \in 0 \ldots \pi$ \label{item:choose2omega}
\item We now have a complete set of elements, and so can SAA as in step \ref{item:SAA} above.\label{item:SAA2}
\item Repeat steps \ref{item:choose2omega} to \ref{item:SAA2} until all $i$ chosen.
\end{enumerate}

\item Repeat the above steps until the parameters $\omega$, $\tau$, $a$, and $e$ are sufficiently sampled for each starting point.

\end{enumerate}

\afterpage{\clearpage}

\subsection{De-orbiting: Unique Sequences of Whole-Pixel Shifts}
The algorithm just described will produce a large number of trial orbits, many of which will be very similar.  The information content of our image is set by the resolution of the telescope, so we can take advantage of this similarity to greatly reduce the number of statistical trials.  This is done by grouping similar orbits into sequences of whole-pixel shift sequences, where the pixels are at least as small as $\mbox{FWHM}/2$.  As we will see, we typically will want to oversample, to say $\mbox{FWHM}/3$, to ensure adequate $SNR$ recovery.

\begin{figure}
\begin{center}
\includegraphics[scale=.92]{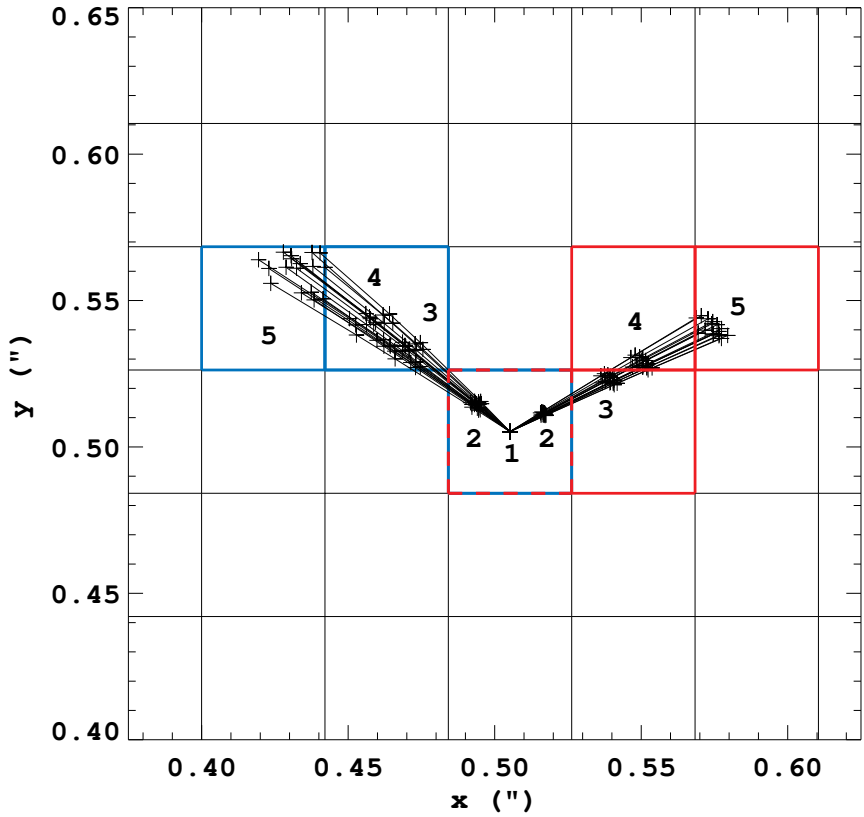}
\end{center}
\caption{Two sequences of whole pixel-shifts, one in red and one in blue. We also show a few of the many orbits that produce these shift sequences.  Once these shifts are determined, a set of 5 images can be de-orbited by shifting the images by the indicated sequence 5-4-3-2-1, that is the pixel containing the orbit in image 2 is is shifted and added to the pixel containing the orbit in image 1, and likewise for images 3, 4, and 5.  Of course, the entire image is shifted, not just single pixels.
\label{fig:pixelation}}
\end{figure}

We calculate the pixel-shift sequence for each orbit by determining which pixel the trial planet (or rather, the center of its PSF) lands on at each time step.  Many orbits end up producing the same sequences of pixel-shifts, and we will keep only the unique ones for use in de-orbiting the observation.  In Figure \ref{fig:pixelation} we illustrate the outcome of the pixel-shift algorithm, showing two unique sequences and a few of the orbits that produced them.

\afterpage{\clearpage}

\begin{figure}
\begin{center}
\includegraphics[scale=.92]{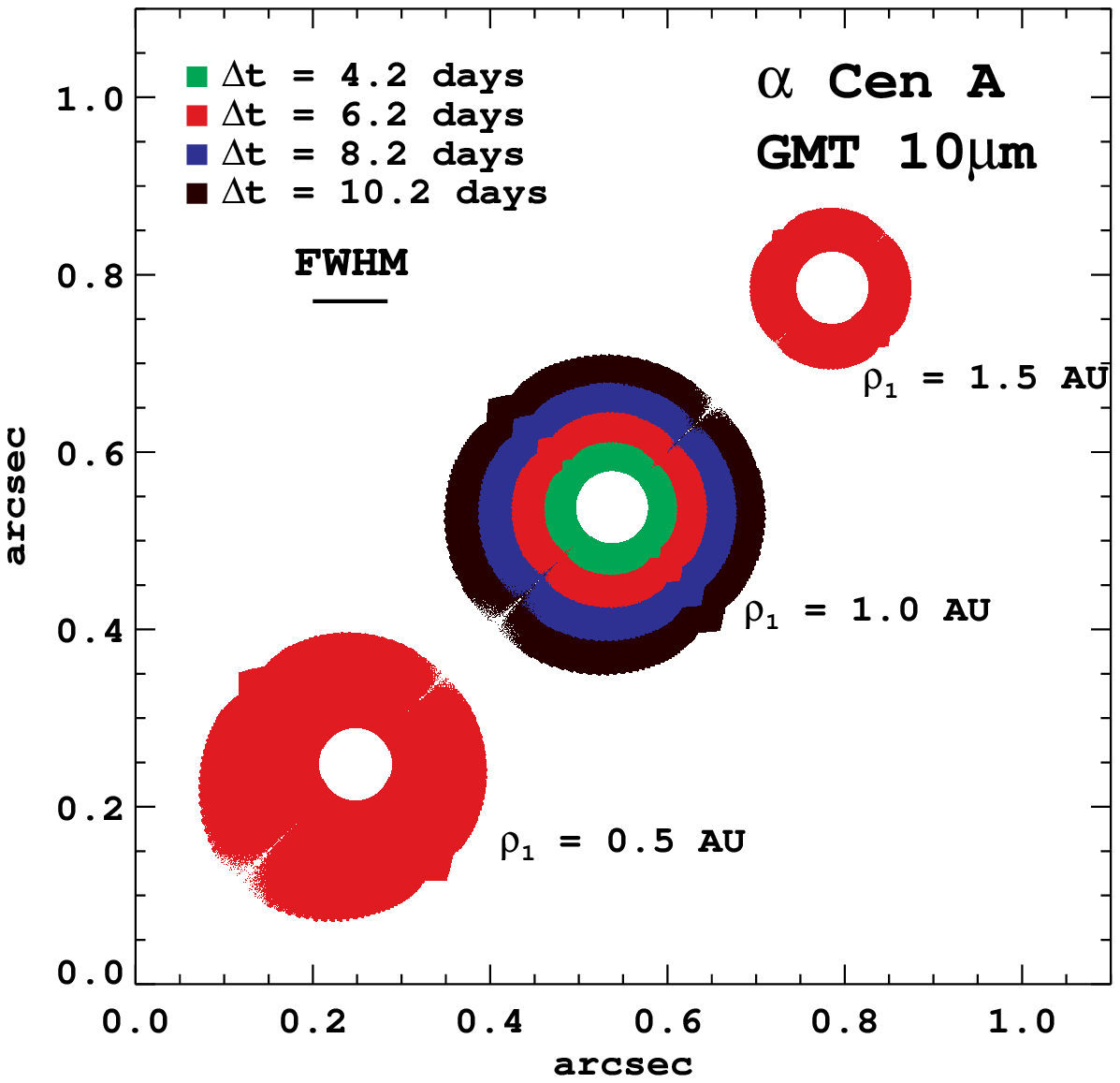}
\end{center}
\caption{Example trial orbits for the GMT, working at $10 \mu$m, observing $\alpha$Cen A.  Plotted are the end points of orbits calculated using the algorithm given in Section \ref{sect:algo} for the given initial projected planet separations $\rho_1$ and elapsed observation times $\Delta t_{tot}$.  The red points show the effect of changing initial separation for a constant elapsed time.  At 1 AU initial separation the colors correspond to different elapsed times as indicated in the legend.  We further analyze these relationships in Table \ref{tab:orbitcalcs} and Figure \ref{fig:props}.  The results of the algorithm appear more complicated than the simple escape-velocity circle analysis in Section \ref{sect:deorbit}.  The end-point clouds are not circularly symmetric about the starting point, and have some azimuthal structure.  For instance there is a triangle extending azimuthally corresponding to face-on high-$e$ orbits, and there are gaps along the radius from the star corresponding to $i$ very near 90$^o$.  These structures are consequences of the chosen grid resolution.  
\label{fig:example_trial_orbits}}
\end{figure}

To test the above algorithm and the pixel-shift technique, we used our GMT $\alpha$ Cen A example and determined the trial orbits for various separations and $\Delta t$s.  We set $\epsilon_{max} = 0.5$ based on our earlier analysis of $SNR$.  The results are summarized in Figure \ref{fig:example_trial_orbits}.  The problem is generally well constrained in that we only have a finite search space for any initial point.  The data used to construct Figure \ref{fig:example_trial_orbits} are provided in Table \ref{tab:orbitcalcs}.  Comparing $N_{orb}$ to $N_{shifts}$, note the large reduction in the number of trials ($\sim10^8$ to $\sim10^2$) due to combining similar orbits.

\afterpage{\clearpage}

\begin{table}
\begin{center}
\begin{tabular}{|c|c|c|c|c|}\hline
$\rho_1$ (AU)&$\Delta t_{tot}$ (days) &No. Obs. & $N_{orb}$ & $N_{shifts}$\\
\hline
0.5& 6.0 & 5 & $2.7\times10^8$& 285\\
\hline
1.0& 2.0 & 5 & $4.1\times10^8$& 14\\
\hline
1.0& 4.0 & 5 & $4.1\times10^8$& 76\\
\hline
1.0& 6.0 & 5 & $4.1\times10^8$& 134\\
\hline
1.0& 8.0 & 5 & $4.1\times10^8$& 253\\
\hline
1.5& 6.0 & 5 & $5.2\times10^8$& 90\\
\hline
1.0& 2.0 & 3 & $4.1\times10^8$& 10\\
\hline
1.0& 4.0 & 5 & $4.1\times10^8$& 78\\
\hline
1.0& 6.0 & 7 & $4.1\times10^8$& 292\\
\hline
1.0& 8.0 & 9 & $4.1\times10^8$& 815\\
\hline

\end{tabular}
\end{center}
\caption{Results of applying the algorithm detailed in Section \ref{sect:algo} for various separations and elapsed observation times.  See also Figure \ref{fig:props}.  Note the dramatic reduction in the number of trials ($N_{orb}$ vs. $N_{shifts}$) after combining similar orbits into whole-pixel shift sequences.
\label{tab:orbitcalcs}
}
\end{table}

\afterpage{\clearpage}

\subsection{$N_{orb}$ Scalings}

\begin{figure}
\begin{center}
\includegraphics[scale=.92]{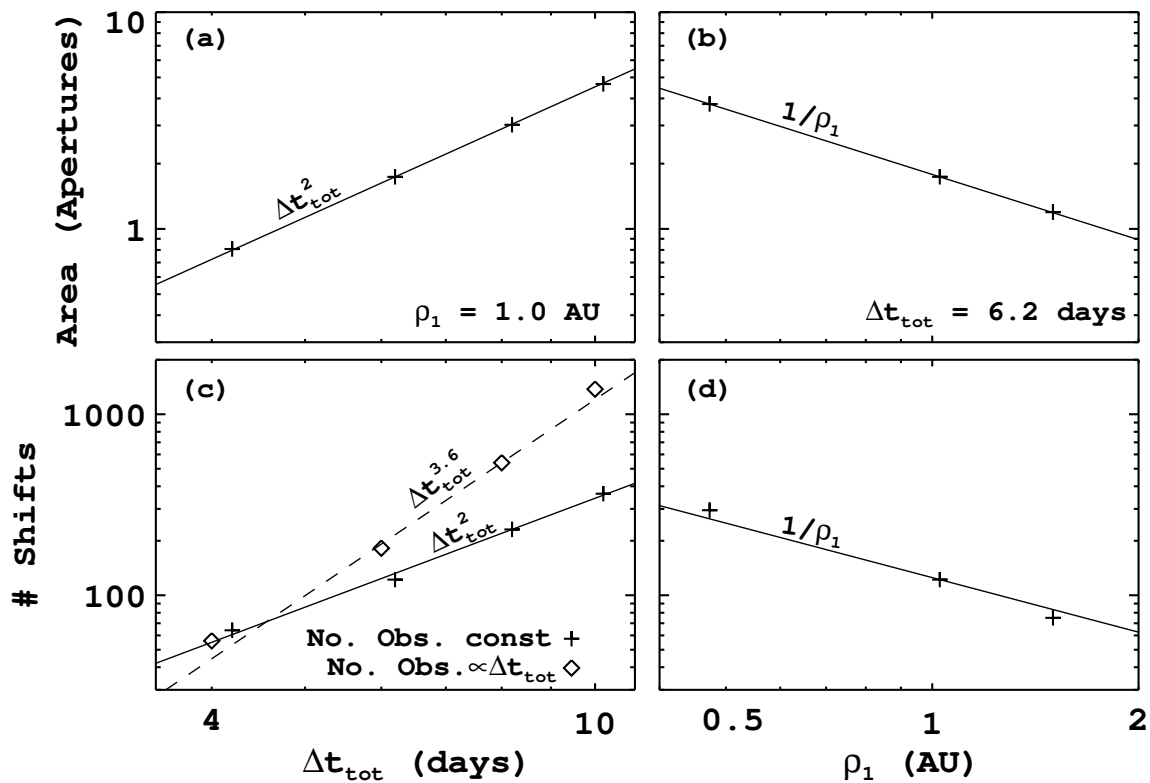}
\end{center}
\caption{Scaling of the number of orbits and the number of resulting whole-pixel shifts with observation elapsed time and with distance from the star.  These results demonstrate that the number of trial orbits $N_{orb}\propto\Delta t_{tot}^4$ scaling that we derived using the escape velocity holds when we rigorously apply orbital mechanics.  Note though that the situation is more complicated with the number of shifts -- if the number of observations increases with elapsed time then the number of shifts grows faster than $\Delta t_{tot}^2$, implying that $N_{orb}$ will increase faster than $\Delta t_{tot}^4$.  These scalings lead to one of our main, if seemingly obvious, conclusions: one must limit the elapsed time of an observation as much as possible when orbital motion is significant.  
\label{fig:props}}
\end{figure}

In Figure \ref{fig:props} we plot the area of the detector which contains the possible trial orbits at $\rho_1 = 1.0$ AU vs. the total elapsed time $\Delta t_{tot}$.  We conclude from this plot that the area around a given starting point is proportional to $\Delta t_{tot}^2$.  Also in Figure \ref{fig:props} we plot area vs separation from the star, and conclude that area is proportional to $1/\rho_1$.  Taken together these results give confidence that the $N_{orb} \propto \Delta t_{tot}^4$ scaling derived earlier holds when we fully apply orbital mechanics rather than the escape velocity approximation.  

Things are a bit more complicated when we consider the scaling of the number of while-pixel shift sequences.  We conducted two sets of trials at $\rho_1 = 1.0$ AU.  In the first, the number of observations and their relative spacing was held constant regardless of $\Delta t_{tot}$.  In the second set, the number of observations scaled with $\Delta t_{tot}$.  As shown in Figure \ref{fig:props}, when the number of observations is constant, the number of shifts scales as $\Delta t_{tot}^2$, but when the number of observations grows with $\Delta t_{tot}$ the number of shifts scales as roughly $\Delta t_{tot}^{3.6}$.  Figure \ref{fig:props}d shows that the number of shifts scales as $1/\rho_1$.  Taken together, we see that for a constant number of observations the pixel-shift technique will follow the $N_{orb} \propto \Delta t_{tot}^4$ scaling. However, if the number of observations also scales with $\Delta t_{tot}$, then our results imply that $N_{orb} \propto \Delta t_{tot}^{5.6}$.  The value of the exponent likely depends on the details of the observation sequence, but this has important implications for observation planning.  

\afterpage{\clearpage}

\subsection{Recovering SNR}

We next consider whether de-orbiting by whole-pixels adequately recovers $SNR$.  To test this we ``orbited'' a Gaussian PSF on face-on orbits with various eccentricities, starting from pericenter.  We then calculated  shifts for detector samplings of 2, 3, and 4 pixels/FWHM, and then de-orbited by these shifts.  The results are summarized in Table \ref{tab:snr_recover}.  On a critically sampled detector we only recover a $5\sigma$ planet to $\sim4.9\sigma$, a $2\%$ loss of $SNR$.  At 3 pixels/FWHM we do much better, recovering $SNR$ to 4.97 for low eccentricities, and 4.95 for higher eccentricities.  Performance for 4 pixels/FHWM sampling is similar.  A 2\% loss of $SNR$ nearly doubles $P_{FA}$, so it appears that we should oversample to at least 3 pixels/FWHM, either optically or by re-sampling images during data reduction.  In our analysis we have assumed that the limiting noise source is background photons (PSF halo or sky), so we ignore the increased readout noise expected from oversampling.

\begin{deluxetable}{cccccccc}
\tablewidth{360pt}
\tablecaption{SNR recovered after de-orbiting with whole-pixel shifts for various samplings.  
\label{tab:snr_recover}}
\startdata
\hline
\hline
Sampling &  \multicolumn{7}{c}{$SNR$ Recovered}\\
(pix/FWHM) &     e=0.0  & e=0.1   &  e=0.2 &    e=0.3   &  e=0.5  &   e=0.7   &   e=0.9\\
\hline
2&            4.89 &   4.89    &   4.89 &     4.88 &    4.86  &     4.86 &      4.86\\
3 &           4.97  &  4.96     &  4.95  &    4.94  &   4.94   &    4.95  &     4.95\\
4  &          4.97   & 4.97      & 4.97   &   4.97   &  4.94    &   4.92   &    4.92\\
\tableline
\enddata
\end{deluxetable}


\subsection{Correlations And The True Impact On $P_{FA}$}

As we have noted several times, the main impact of orbital motion is to reduce $SNR$, which in turn reduces our statistical sensitivity.  If we attempt to de-orbit an observation in order to recover $SNR$, we do so at the cost of a large increase in the number of trials.  Worst case, this results in a proportional increase in $FAR$ since nominally $FAR = P_{FA}\times N_{orb}$.  However, we expect significant correlation between trials of neighboring orbits and whole-pixel shifts.  To investigate this, we performed a series of monte carlo experiments.  A sequence of images with Gaussian noise was generated, and first stacked without shifting, hereafter called the naive-add. The same sequence was then shifted by each possible whole-pixel shift, assuming a 1AU initial separation around $\alpha$ Cen A.  This experiment was conducted for observations with total elapsed times $\Delta t_{tot}$ of 4.2, 6.2, 8.2, and 10.2 days, with samplings of 2, 3, and 4 pixels/FWHM.

We performed several tests on each sequence.  The first was a simple threshold test on the naive-add, with the threshold set for the worst case orbital motion given by Equation \ref{eqn:s_gnd} with $v_{om} = v_{esc}$.  We performed simple aperture photometry, with a $r_{ap}=1$ FHWM.  As expected the resultant $P_{FA1}$ is as predicted by Equation \ref{eqn:gauss_pfa}.  The next test was to apply a $5\sigma$ threshold after de-orbiting by whole-pixel shifts and adding.  If all shifts were completely uncorrelated, then we would expect $FAR = (2.9\times 10^{-7}) \times N_{shifts}$,  but as we predicted, shifts are correlated and  $P_{FA2}$ is lower than this.  

The final test performed was to apply both thresholds in sequence, such that a detection is made only if the naive-add results in $SNR$ greater than the threshold for worst case orbital motion, {\bf and} the de-orbited SAA results in $SNR$ $>5$.  This $P_{FA3}$ is lower than either $P_{FA1}$ or $P_{FA2}$, but still higher than if no orbital motion occurred.  

The results of each trial are present in Table \ref{tab:mc_corr}.  Applying both threshold tests results in significant improvement over the naive-add in terms of $FAR$.  Another interesting result is that sampling has only a minor impact on $P_{FA3}$.  This makes some sense as we expect the correlation of neighboring shifts to be set by the FWHM, not the sampling.  So even though the accuracy of $SNR$ recovery is improved, and quite a few more shifts are required, these shifts remain correlated across the same spatial scale resulting in little change in the overall $FAR$.

\begin{deluxetable}{cccccc}
\tablewidth{350pt}
\tablecaption{False alarm probabilities after de-orbiting Gaussian noise images.  
\label{tab:mc_corr}}
\startdata
\hline
\hline
$\Delta t_{tot}$ (days)\tablenotemark{1} &  $SNR_t$\tablenotemark{2} &   $N_{shifts}$\tablenotemark{3} &  $P_{FA1}$\tablenotemark{4} &    $P_{FA2}$\tablenotemark{5} &      $P_{FA3}$\tablenotemark{6}\\
\hline
\cutinhead{2 pixels/FWHM}
\hline
4.2  &  4.635 &  64  &   $1.74\times 10^{-6}$ &  $7.65\times 10^{-6}$ &   $8.06\times 10^{-7}$\\
6.2  &  4.220 & 122  &   $1.24\times 10^{-5}$ &  $1.52\times 10^{-5}$ &   $2.70\times 10^{-6}$\\
8.2  &  3.330 & 231  &   $4.40\times 10^{-4}$ &  $2.71\times 10^{-5}$ &   $9.93\times 10^{-6}$\\
10.2 &  2.625 & 364  &   $4.33\times 10^{-3}$ &  $4.03\times 10^{-5}$ &   $2.17\times 10^{-5}$\\
\tableline
\cutinhead{3 pixels/FWHM}
\tableline
4.2  & 4.635 &   108  &  $2.11\times 10^{-6}$ & $1.37\times 10^{-5}$  & $4.80\times 10^{-7}$\\
6.2  & 4.220 &   285  &  $1.21\times 10^{-5}$ & $3.39\times 10^{-5}$  & $2.04\times 10^{-6}$\\
8.2  & 3.330 &   496  &  $4.31\times 10^{-4}$ & $5.64\times 10^{-5}$  & $9.96\times 10^{-6}$\\
10.2 & 2.625 &   741  &  $4.34\times 10^{-3}$ & $8.15\times 10^{-5}$  & $2.67\times 10^{-5}$\\
\tableline
\cutinhead{4 pixels/FWHM}
\tableline
4.2  & 4.635 &  217   &  $1.78\times 10^{-6}$ & $2.64\times 10^{-5}$  & $4.44\times 10^{-7}$\\
6.2  & 4.220 &  487   &  $1.24\times 10^{-5}$ & $5.61\times 10^{-5}$  & $1.48\times 10^{-6}$\\
8.2  & 3.330 &  844   &  $4.35\times 10^{-4}$ & $9.19\times 10^{-5}$  & $1.14\times 10^{-5}$\\
10.2 & 2.625 & 1315   &  $4.32\times 10^{-3}$ & $1.41\times 10^{-4}$  & $3.15\times 10^{-5}$\\
\tableline
\enddata
\tablenotetext{1}{Elapsed time of the observation.}
\tablenotetext{2}{$SNR$ threshold from Equation \ref{eqn:s_gnd}, using $v_{orb} = \sqrt{2}v_{FOC}.$}
\tablenotetext{3}{Number of unique whole-pixel shifts required to de-orbit.}
\tablenotetext{4}{False alarm probability for the naive-add, from MC experiment results.  Expected values given by Equation \ref{eqn:gauss_pfa}.}
\tablenotetext{5}{False alarm probability after de-orbiting with whole-pixel shifts.}
\tablenotetext{6}{False alarm probability after testing both the naive-add and de-orbiting.}
\end{deluxetable}

\afterpage{\clearpage}

\subsection{Impact on Completeness of the Double Test}
There is \emph{still} an impact on completeness, however, because we are now conducting two trials instead of one.  This lowers the true positive probability ($P_{TP}$).  Consider a $5\sigma$ planet on the worst case fastest possible orbit, for the 10.2 day elapsed time case.  The threshold for the naive add is 2.625.  We have a 50\% probability of detecting this planet after the naive add.  If it is detected on the first test, there is then some probability $P_{TP} < 1$ of detecting at $SNR \ge 5$ after de-orbiting.  Worst case, this will be 50\%, resulting in a net $P_{TP}$ of 25\%.  In reality, it will be better than this as the two trials will be strongly correlated.

Even if this worst case of 25\% were realized this is still significant improvement over Option I.  A $2.6\sigma$ signal would only be detected 10\% of the time with a $5\sigma$ threshold.  Given the reduction in $P_{FA}$ from $4.3\times10^{-3}$ to $2.2\times10^{-5}$, likewise an improvement over Option II at $2.6\sigma$, \emph{it is clear that de-orbiting by whole-pixel shifts does improve our ability to detect an orbiting planet}.  The situation will be even better for slower planets, and most of the area searched will not be subject to the worst case orbital speed.  We leave a complete analysis of the impact on search completeness for future work.  One can also imagine adjusting the thresholds to optimize completeness at the expense of worse $P_{FA}$.

\subsection{Tractability of a Blind Search}

We end this section by concluding that a blind search when orbital motion is significant is tractable.  Orbital motion will make such a search less sensitive, both in terms of number of false alarms and in terms of completeness, but Keplerian mechanics gives us enough tools to bound the problem. As we have shown de-orbiting a sequence of observations can recover $SNR$ to its nominal value, and we can do so while controlling the impact on statistical sensitivity.  For the $\Delta t_{tot} = 6.2$ day observation, $P_{FA3}$ was roughly a factor of 10 higher than if no orbital motion occurred.  This increase only occurs over a bounded region around the star, so the net effect on $FAR$ will be contained.  Using this factor of 10 as the mean value over the 7.7 AU = 69.1 FWHM radius region around $\alpha$ Cen A where orbital motion is significant, the $FAR$ in this area will have gone from  $\sim1/1000$ to $\sim1/100$ in our GMT/10$\mu$m example.  The key, though, appears to be to limit the elapsed time of the observation as the number of trials increases --- decreasing sensitivity -- proportionally to \emph{at least} $\Delta t_{tot}^4$ in a blind search.

The main caveat at this point in our analysis is that we have drawn the conclusion of tractability using Gaussian statistics.  It is well known that speckle noise, which will often be the limiting noise source for high contrast imaging in the HZ, is not Gaussian and results in much higher $P_{FA}$ for a given SNR \citep{2008ApJ...673..647M}.  Future work on this problem will need to take this into account.

Next we consider a more strongly bounded scenario, where we have significant prior information about the orbit of the planet from radial velocity surveys.

\afterpage{\clearpage}

\section{Cued Search: Using RV Priors}
\label{sec:cued}

The situation is greatly improved if we have prior information, such as orbit parameters from RV or astrometry.  Here we consider the case of Gliese 581d, and the previously discussed future observation of this planet by EPICS at the E-ELT \citep{2010SPIE.7735E..81K}.  There is some controversy surrounding the solution to the RV signal, and whether planet d even exists \citep{2011arXiv1109.2505F, 2012arXiv1207.4515V, 2012arXiv1209.3154B}.  We show results for both the floating eccentricity Keplerian fits of \citet{2011arXiv1109.2505F}[hereafter F11], and the all circular interacting model of \citet{2012arXiv1207.4515V}[hereafter V12].  Doing so allows us to illustrate the impact of eccentricity on the analysis, and prevents us having to take a stand in a currently raging debate.  The parameters used herein are listed in Table \ref{tab:orbparams}.

\begin{table}
\begin{center}
\begin{tabular}{|l|c|c|c|c|c|c|c|}\hline
Model                      &       $a$ (AU)&      $e$     &  $\omega$ (deg) & $\sigma_{t_0}$ (days)\\
\hline
\citet{2011arXiv1109.2505F}&$0.218 \pm 0.005$ & $0.25 \pm 0.09$ & $356.0 \pm 19.0$ & $\pm 3.4$\\
\citet{2012arXiv1207.4515V}&$0.218 \pm 0.005$ & $0.0 \pm 0.0$ & $0.0 \pm 0.0$ & $\pm7.45$ \\
\hline
\end{tabular}
\end{center}
\caption{Orbital parameters for Gl 581d used in this analysis.  We derived the values reported here from other parameters where necessary.  Only the uncertainty in $t_0$ impacts our analysis.  In both models the orbital period is $66.6$ days.
\label{tab:orbparams}
}
\end{table}

\afterpage{\clearpage}

Instead of a grid search, we use a monte carlo (MC) method.  The RV technique provides the parameters $a$, $e$, $\omega$ and $t_0$ or their equivalents.  We can take the results of fitting orbits to the RV signal, and the associated uncertainties, as prior distributions which we sample to form trial orbits.  We will assume that all uncertainties are uncorrelated and are from Gaussian distributions.  

We assume that the 20 hr integration is broken up over 6.2 nights based on the same logic discussed in Section \ref{sec:blind}.  \cite{2010SPIE.7735E..81K} actually assumed $20\times1$ hr observations based on the amount of rotation needed, but did not consider the effects of orbital motion over 20 days of a 67 day period (M. Kasper, personal communication (2012)).  

\subsection{Constraints}
In order to minimize the number of trial orbits to consider, we can apply various constraints taking advantage of the information we have from the RV detection.

In the case of a multi-planet system dynamical analysis can place constraints on the inclination based on system stability.  For Gl 581, \citet{2009A&A...507..487M} found the system was stable for $i > 30$.  We can also make use of the geometric prior for inclination, where we expect $P_i = \sin(i)$ in a population of randomly oriented systems.

Since this is a reflected light observation, the orbital phase and its impact on the brightness of the planet must be considered.  The planet's reflected flux is given by
\begin{equation}
F_p(\alpha) = F_*\left(\frac{R_p}{r}\right)^2 A_g(\lambda) \Phi(\alpha)
\end{equation}
where $F_*$ is the stellar flux, $R_p$ is the planet's radius, $r$ its separation, $A_g(\lambda)$ is the wavelength dependent geometric albedo,  and $\Phi$ is the phase function at phase angle $\alpha$. The phase angle is given by
\begin{equation}
\cos(\alpha) = \sin(f + \omega)\sin(i).
\end{equation}
In general, determining the quantity $A_g(\lambda)\Phi(\alpha)$ requires atmospheric modeling \citep{2010ApJ...724..189C}.  For now, we assume that $\Phi$ follows the Lambert phase function
\begin{equation}
\Phi(\alpha) = \frac{1}{\pi}\left[\sin(\alpha) + (\pi - \alpha)\cos(\alpha)\right]
\end{equation}
We assume that the prediction of \citet{2010SPIE.7735E..81K} was made for the planet at quadrature, $\alpha = \pi/2$, where $\Phi = 0.318.$  We then require that the mean value of $\Phi$ during the observation be greater than this value - that is the planet is as bright or brighter than it is at quadrature.

\subsection{Initial Detection}

An important consideration in an RV-cued observation will be when to begin.  As a first approximation, we assume that maximizing planet-star separation will maximize our sensitivity. This may not be true when working in reflected light due to the phase and separation dependent brightness of the planet in this regime.  Proceeding with the approximation for now, we expect to plan this observation to be as close to apocenter as possible.  In this case we will begin integrating 3.1 days before $t_0+P/2$.  

\begin{figure}
\begin{center}
\includegraphics[scale=.7]{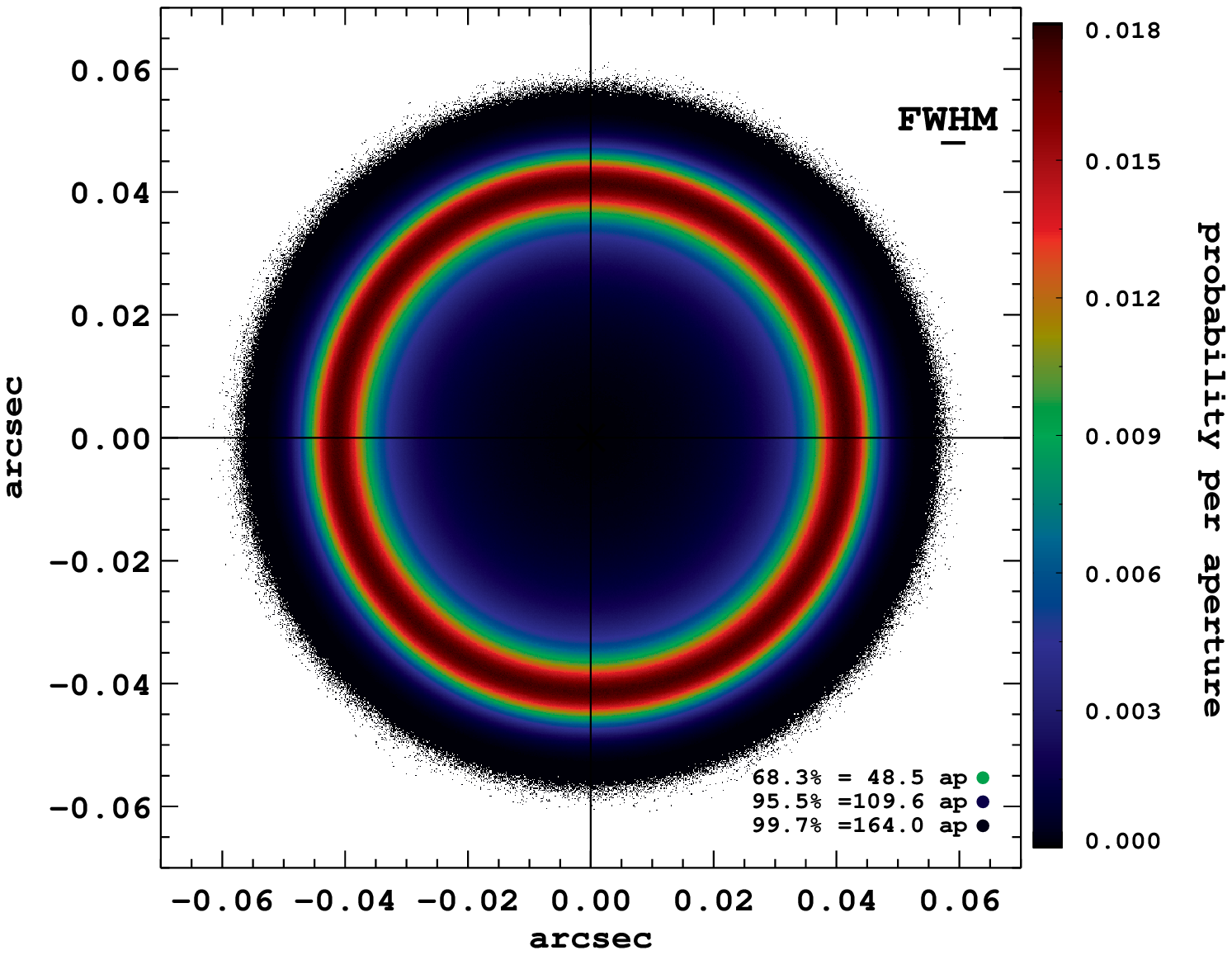}
\includegraphics[scale=.7]{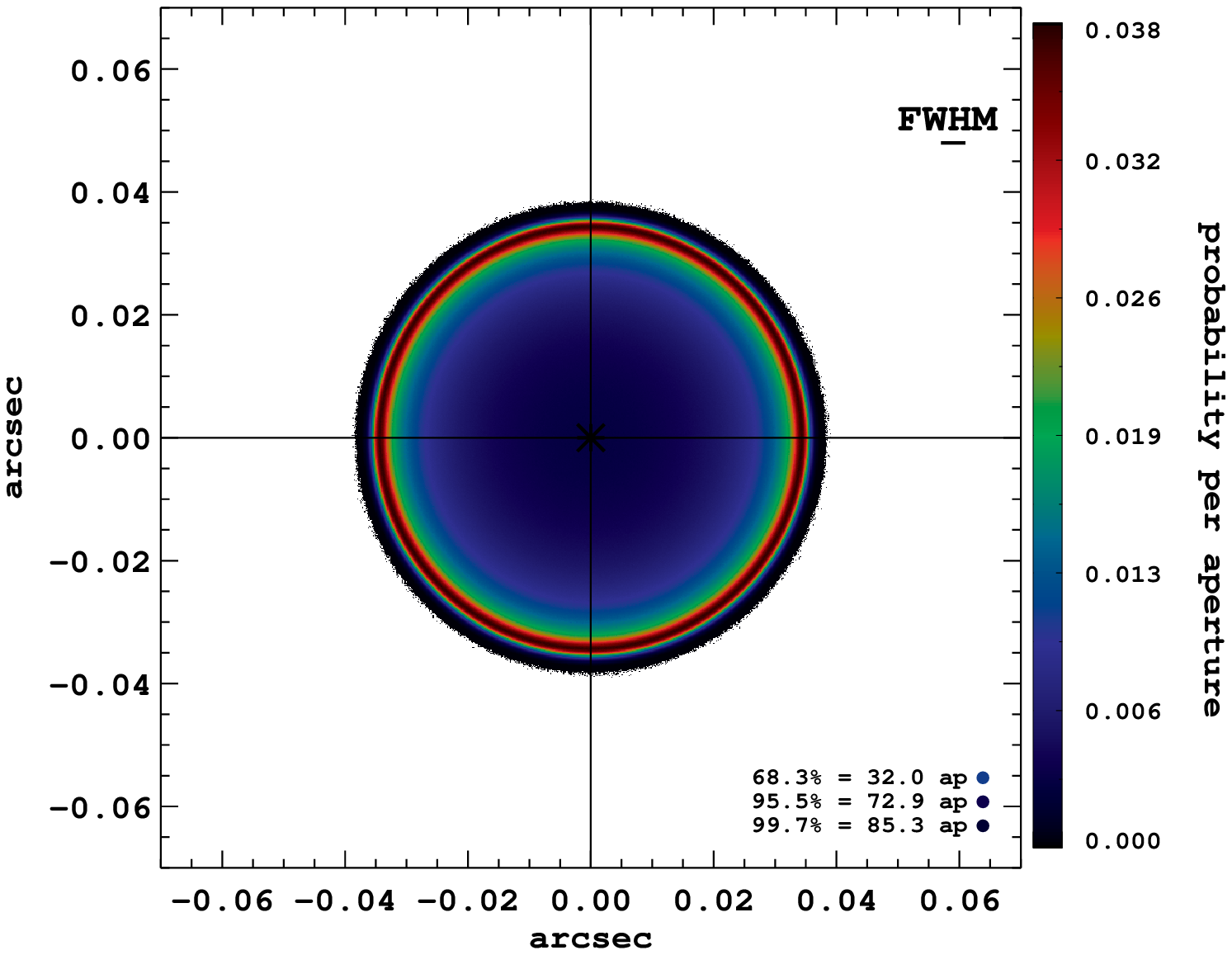}
\end{center}
\caption{Possible starting points for Gl 581d, observed near apocenter.   Top: using the parameters of \citet{2011arXiv1109.2505F}'s eccentric model.  Bottom: assuming the parameters of \citet{2012arXiv1207.4515V}'s circular interacting model.   The color shading is in units of probability per aperture (each aperture has area $\pi\mbox{FWHM}^2$).  The legend indicates the color which encloses the given completeness intervals, and the enclosed area in apertures, which can be directly related to the false alarm rate as discussed in the text.
\label{fig:rvic}}
\end{figure}

To understand the area where we will be searching for Gl 581d, we first conducted an MC experiment to calculate the possible positions of the planet at $t=t_0+P/2-3.1$ days.  To do so, we drew random values of $a$, $e$, $w$, and $t_0$ from Gaussian distributions with the parameters of Table \ref{tab:orbparams}.  We drew a random value of $i$ from the $\sin(i)$ distribution, and rejected any value of $i \leq 30$ based on the dynamical prior.  Finally $\Omega$ was drawn from a uniform distribution in $0\ldots2\pi$.  This process was repeated $10^9$ times, and the frequency at which starting points occur in the area around the star was recorded.  The results are shown in Figure \ref{fig:rvic} for the V12 circular model and for the F11 eccentric model.  The Figure shows the area which must be searched to obtain various completeness.  For instance, if we desire 95\% completeness in the V12 model, we must consider an area of 71 apertures.  Since this $SNR=5$ detection is broken up into 5 distinct integrations, our first attempt will have $SNR = 2.24$, giving a $FAR = 0.89$ for the first 4 hr integration.  In other words, we should expect a false alarm in addition to a real detection.

\afterpage{\clearpage}

\subsection{Calculating Orbits and Shifts}

Now we assume that we have an initial detection at $SNR \sim 2.24$ within the highest probability regions\footnote{For the purposes of this analysis, we calculated initial separation $\rho_1$ using the mean parameters for each model and an inclination $i=60$}.  In order to follow-up this detection over subsequent nights, we must determine the possible locations of the planet, constrained by the RV-derived orbital elements.  

\begin{figure}
\begin{center}
\includegraphics[scale=.645]{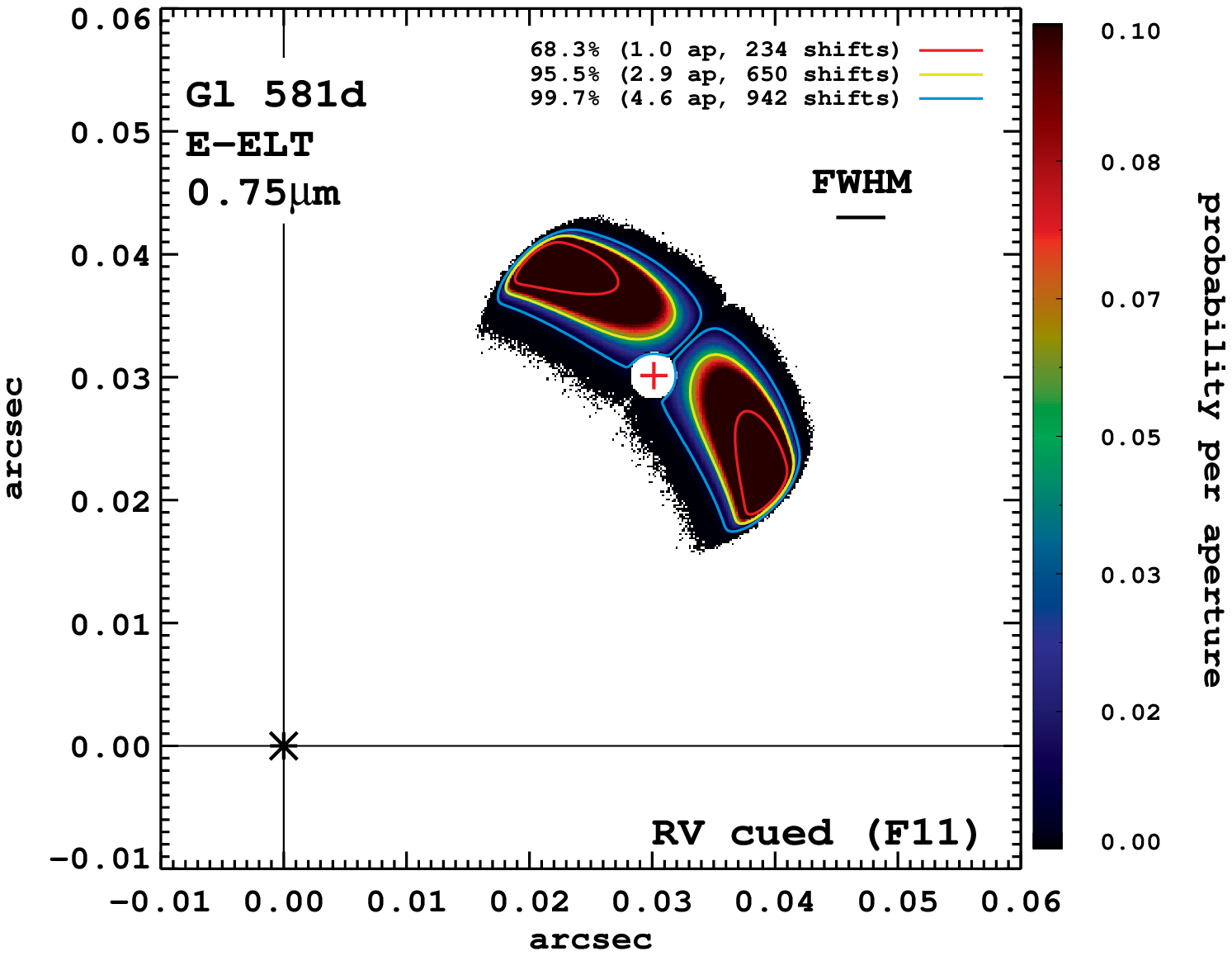}
\includegraphics[scale=.645]{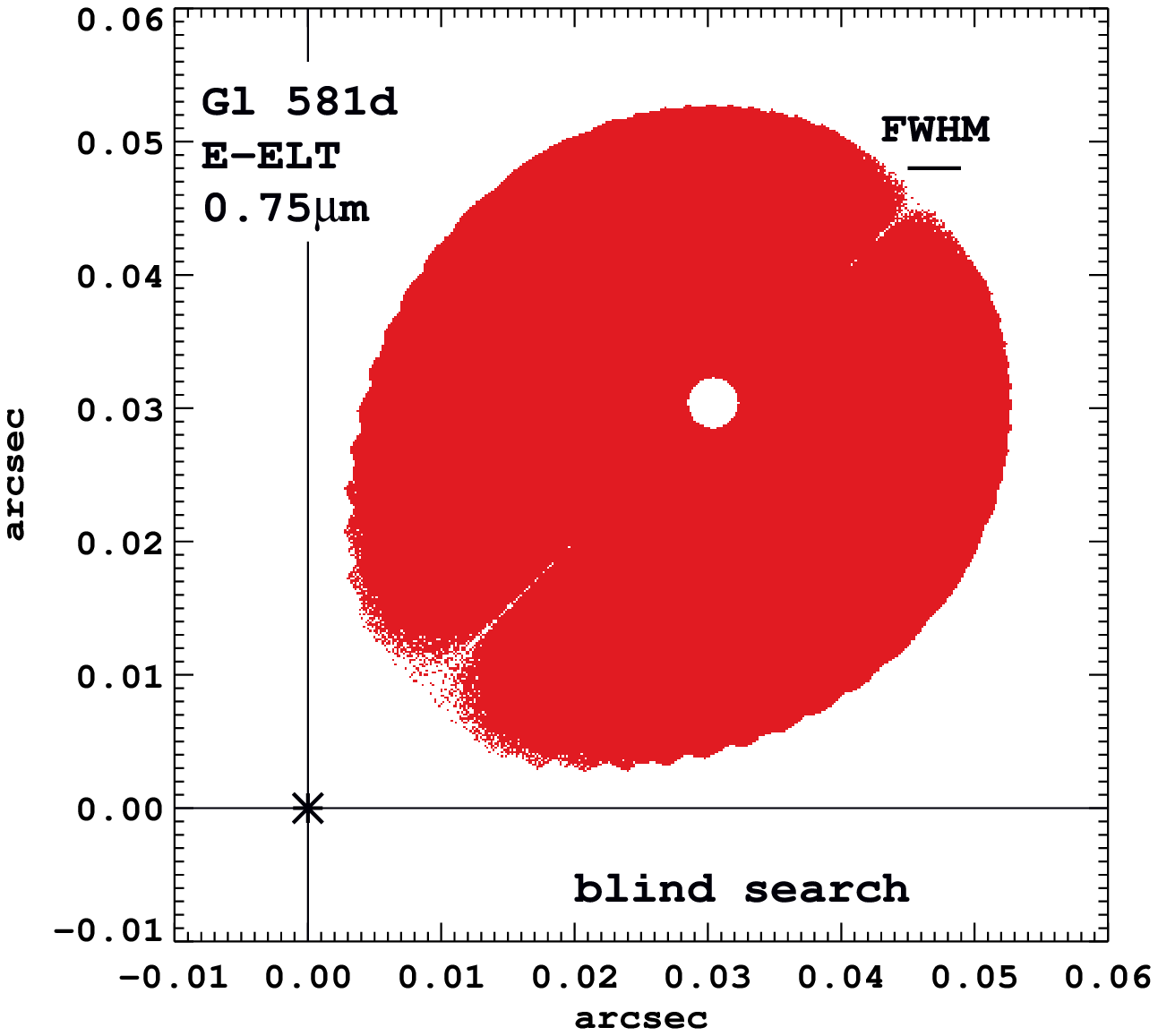}
\end{center}
\caption{Trial orbits for Gl 581d, observed near maximum elongation.  In the top panel we use the parameters of \citet{2011arXiv1109.2505F}'s eccentric model.  The bottom panel shows the results for a blind search from the same starting point.   The red cross shows the starting point, and the star is located at the origin.  The top panel color shading is in units of probability per aperture (each aperture has area $\pi\mbox{FWHM}^2$).  The legend indicates the color which encloses the given completeness intervals, the enclosed area in apertures, and the number of unique whole-pixel shift sequences which must be tried in order to de-orbit the observation. The number of shift sequences is directly related to the false alarm rate, and hence the sensitivity.  For comparison, the blind search algorithm produced $\sim 12000$ shifts.  RV cueing greatly improves our sensitivity in the presence of orbital motion.
\label{fig:gl581_cloudcomp}}
\end{figure}

We proceed by choosing $a$, $e$, $\omega$, and $t_0$ from Gaussian distributions as above.  Now as long as $r > \rho$ we will have a unique solution for $i$ and $\Omega$ given the randomly chosen parameters (see the blind search algorithm above).  We take into account dynamical stability by rejecting any orbit which has $i \le 30$.  The orbit determined in this fashion was then projected 6.2 days into the future and the frequency of these final points was recorded.  We show the result for the V11 model in Figure \ref{fig:gl581_cloudcomp}, top panel.  Using the RV determined parameters and their uncertainties allows us to determine the probability density of orbit endpoints, and determine how much of the search space we must consider for a given completeness.  The whole-pixel shifts were also calculated using a sampling of FWHM/3, and are shown in the legend. We also applied the blind search algorithm to this observation from the same starting point, and show the results for comparison in the bottom panel of  Figure \ref{fig:gl581_cloudcomp}.  As expected the RV priors significantly reduce the search space - we have 942 trial shift-sequences to consider instead of 12000.  

\afterpage{\clearpage}

\begin{figure}
\begin{center}
\includegraphics[scale=.66]{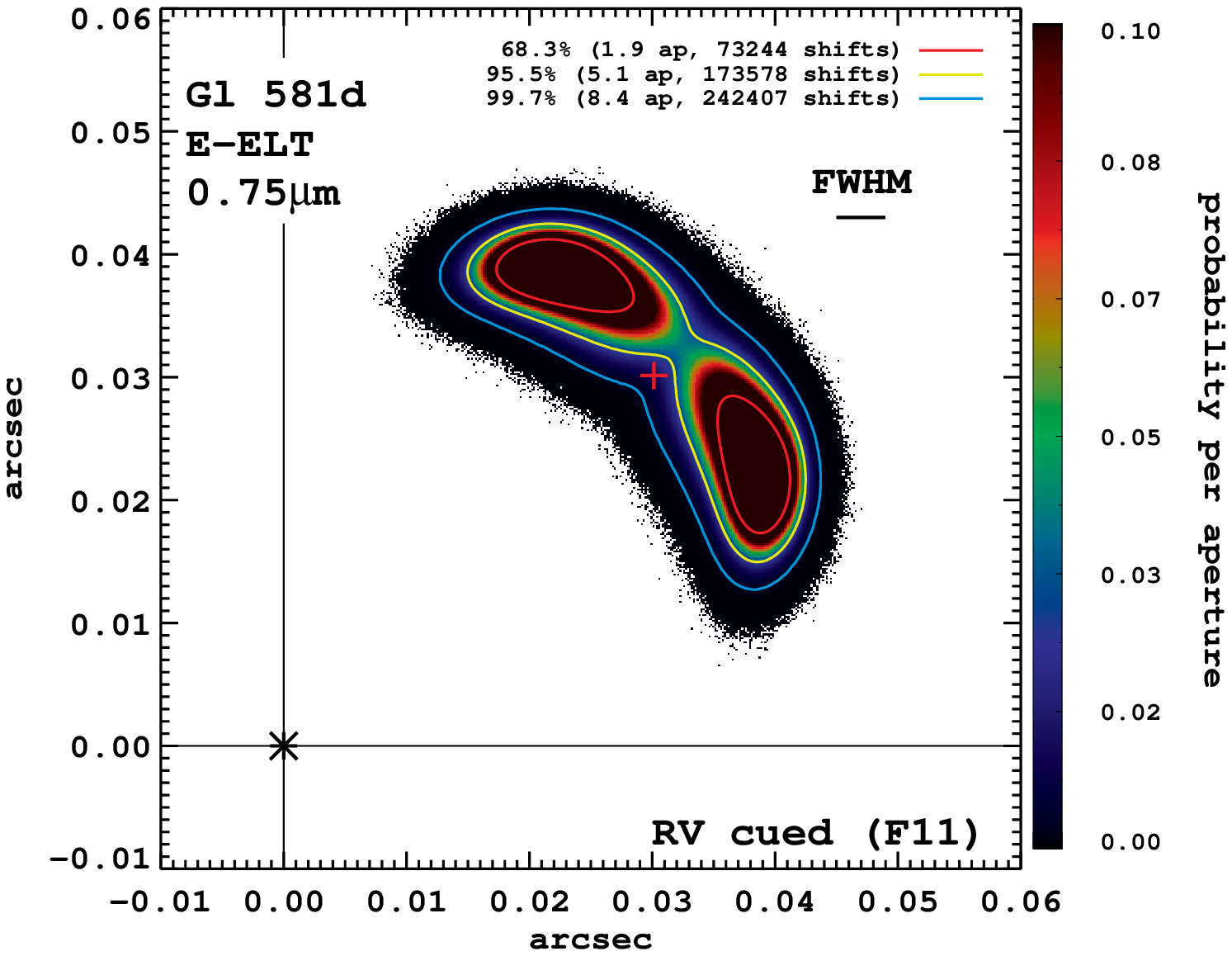}
\includegraphics[scale=.66]{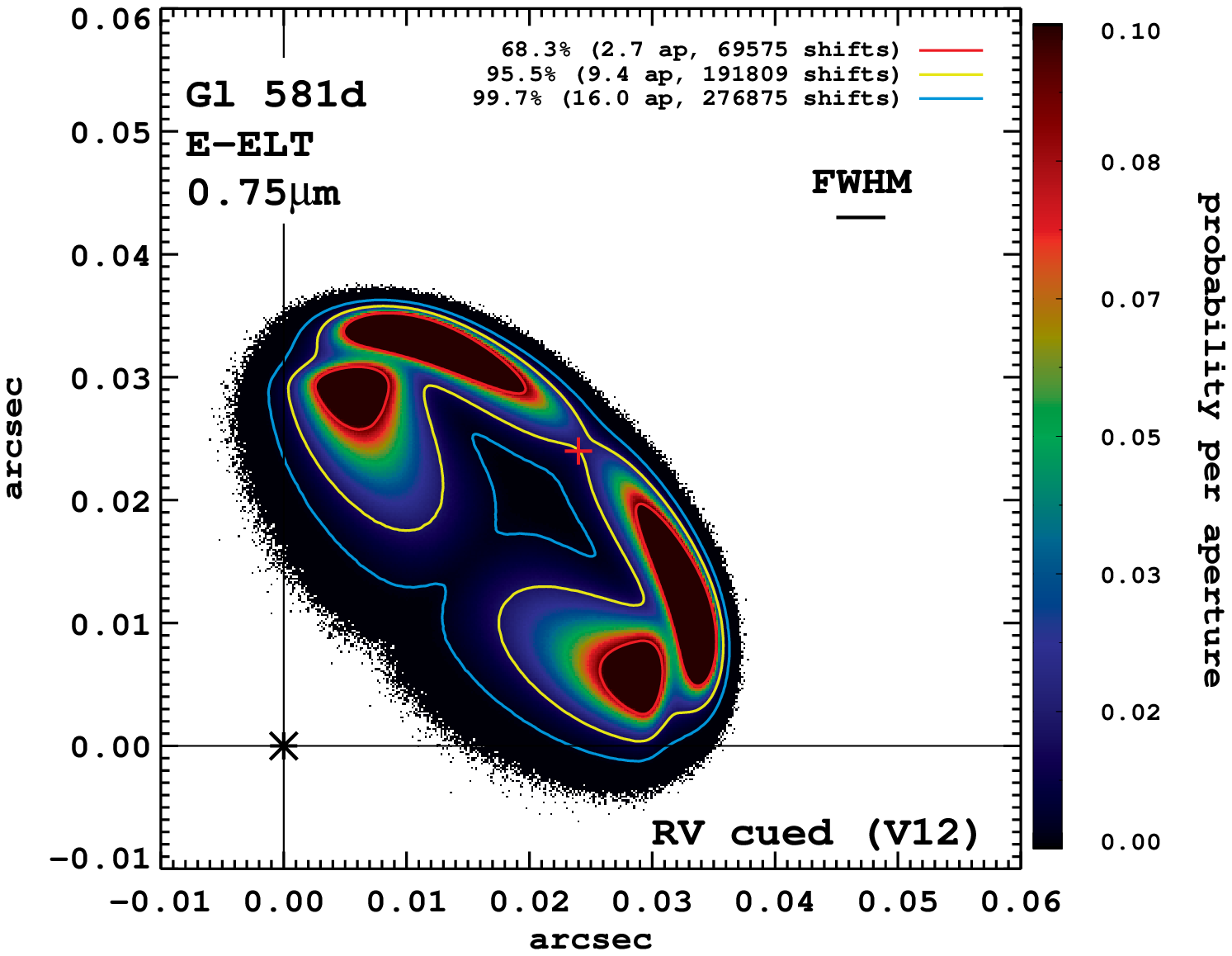}
\end{center}
\caption{Trial orbits for Gl 581d, observed near maximum elongation, assuming the parameters of (top)  \citet{2011arXiv1109.2505F}'s Keplerian eccentric model and (bottom) \citet{2012arXiv1207.4515V}'s circular interacting model.  In this simulation we allowed the initial position to vary with standard deviation $\sigma_{x,y} = FWHM/SNR$.  The red cross shows the starting point, and the star is located at the origin.  The color shading is in units of probability per aperture (each aperture has area $\pi\mbox{FWHM}^2$).  The legend indicates the color which encloses the given completeness intervals, the enclosed area in apertures, and the number of unique whole-pixel shift sequences which must be tried in order to de-orbit the observation. The number of shift sequences is directly related to the false alarm rate, and hence the sensitivity.
\label{fig:gl581_rv}}
\end{figure}

Another important consideration here is that our initial $2.24\sigma$ detection will have a large position uncertainty,  which we estimate by $\sigma_{\rho_0} = FWHM/SNR$.  We added a random draw for the starting position, and repeated the MC experiment for F11 and also conducted a run for the V12 parameters.  The results are shown in Figure \ref{fig:gl581_rv}. The number of shift sequences is much higher due to the uncertainty in the starting position caused by our low $SNR$ initial detection, but we expect correlations to come to the rescue as in our $\alpha$ Cen example.  To compare to Figure \ref{fig:gl581_cloudcomp} keep in mind that the blind search would have to be applied to all $~5500$ pixels in the search space indicated by Figure \ref{fig:rvic}. 

As in the GMT/$\alpha$ Cen example, we leave for future work a complete analysis of sensitivity and completeness.  The large number of trial shifts calculated when we include uncertainty in the starting position motivates us to suggest that we will ultimately turn this analysis over to a much more robust optimization strategy, such as a Markov Chain Monte Carlo (MCMC) routine.  Once an area of the image was identified with a high post-shift $SNR$, a MCMC analysis could determine the very best orbit and assign robust measures of significance to the result.  

We also note that these results likely overestimate the number of trial orbits since we have assumed uncorrelated errors.  In reality the RV best fit parameters are likely strongly correlated, which should act to reduce the number of orbits to consider.  

\afterpage{\clearpage}

\section{Conclusions}
\label{sec:conclusions}
In the coming campaigns to directly image planets in the HZs of nearby stars, orbital motion will be large enough to degrade our sensitivity.  This effect has been ignorable in direct imaging campaigns to date, which have typically looked for wide separation planets.  We have analyzed this issue in some detail, and shown that applying basic Keplerian orbital mechanics allows us to bound the problem sufficiently that we believe direct imaging in the HZ to be a tractable problem.  Our main conclusions are:

(1) When projected onto the focal plane, a planet in a face-on circular orbit moves with speed given by
\begin{equation}
v_{FOC} = 0.0834\left(\frac{D}{\lambda d}\right)  \sqrt{\frac{M_*}{a}} \mbox{ FWHM day}^{-1}.
\end{equation}
In the HZ of nearby stars, especially when considering giant telescopes, speeds are high enough that planets will move significant fractions of a PSF FWHM during a single observation.  This smears out the planet's flux resulting in a lower $SNR$.

(2) In background limited photometry, an $SNR$ maximum is reached after about $\sim 2$ FWHM of motion has occurred on the focal plane.  From there, integrating longer offers no improvement with a fixed-size aperture.  Adapting the aperture could mitigate this to some extent, but at the cost of significantly longer exposure times.

(3) When $SNR$ is reduced by orbital motion, we have three options.  Option I is to do nothing, and accept the loss of completeness due to planets appearing fainter.  Option II is to adjust our detection threshold at the cost of more false alarm detections.  Option III is to de-orbit an observation, recovering $SNR$ to its nominal value, but also at the cost of more false alarms.

(4) For exposure times of 10s of hours, we expect an observation to extend over several days under realistic assumptions about ground based observing.  If we naively attempt to de-orbit such an observation, the false alarm rate per star will increase by at least $FAR\propto\Delta t_{tot}^4$, where $\Delta t_{tot}$ is the total elapsed time of the observation.  

(5) De-orbiting a sequence of shorter exposures is possible, and tractable.  Taking advantage of strong correlations between trial orbits, we will realize increases in the $FAR$ on the order of a factor of 10 in the region around a star where orbital motion matters.  Since this will be a small, bounded region, this increase in $FAR$ appears to be acceptable.

(6) Cueing from another detection method, such as RV, provides significant benefit.  It allows us to initiate our search at the optimum time, and significantly reduces the size of the search space.  Having prior distributions for some of the orbital elements will allow us to efficiently determine where and how to search to optimize completeness.

We thank the anonymous referee for insightful and constructive comments.  We thank Jessica Orwig for reviewing this manuscript.  JRM is grateful for the generous support of the Phoenix ARCS foundation.  LMC and JRM acknowledge support from the NSF AAG.

\bibliographystyle{apj}
\bibliography{ms}

\begin{thebibliography}{42}
\expandafter\ifx\csname natexlab\endcsname\relax\def\natexlab#1{#1}\fi

\bibitem[{{Baluev}(2012)}]{2012arXiv1209.3154B}
{Baluev}, R.~V. 2012, ArXiv e-prints

\bibitem[{{Bernstein} {et~al.}(2004){Bernstein}, {Trilling}, {Allen}, {Brown},
  {Holman}, \& {Malhotra}}]{2004AJ....128.1364B}
{Bernstein}, G.~M., {Trilling}, D.~E., {Allen}, R.~L., {Brown}, M.~E.,
  {Holman}, M., \& {Malhotra}, R. 2004, \aj, 128, 1364

\bibitem[{{Biller} {et~al.}(2007){Biller}, {Close}, {Masciadri}, {Nielsen},
  {Lenzen}, {Brandner}, {McCarthy}, {Hartung}, {Kellner}, {Mamajek}, {Henning},
  {Miller}, {Kenworthy}, \& {Kulesa}}]{2007ApJS..173..143B}
{Biller}, B.~A., {et~al.} 2007, \apjs, 173, 143

\bibitem[{{Brown}(2004)}]{2004ApJ...607.1003B}
{Brown}, R.~A. 2004, \apj, 607, 1003

\bibitem[{{Brown}(2005)}]{2005ApJ...624.1010B}
---. 2005, \apj, 624, 1010

\bibitem[{{Brown} \& {Soummer}(2010)}]{2010ApJ...715..122B}
{Brown}, R.~A., \& {Soummer}, R. 2010, \apj, 715, 122

\bibitem[{{Bruntt} {et~al.}(2010){Bruntt}, {Bedding}, {Quirion}, {Lo Curto},
  {Carrier}, {Smalley}, {Dall}, {Arentoft}, {Bazot}, \&
  {Butler}}]{2010MNRAS.405.1907B}
{Bruntt}, H., {et~al.} 2010, \mnras, 405, 1907

\bibitem[{{Cahoy} {et~al.}(2010){Cahoy}, {Marley}, \&
  {Fortney}}]{2010ApJ...724..189C}
{Cahoy}, K.~L., {Marley}, M.~S., \& {Fortney}, J.~J. 2010, \apj, 724, 189

\bibitem[{{Cavarroc} {et~al.}(2006){Cavarroc}, {Boccaletti}, {Baudoz}, {Fusco},
  \& {Rouan}}]{2006A&A...447..397C}
{Cavarroc}, C., {Boccaletti}, A., {Baudoz}, P., {Fusco}, T., \& {Rouan}, D.
  2006, \aap, 447, 397

\bibitem[{{Chauvin} {et~al.}(2012){Chauvin}, {Lagrange}, {Beust}, {Bonnefoy},
  {Boccaletti}, {Apai}, {Allard}, {Ehrenreich}, {Girard}, {Mouillet}, \&
  {Rouan}}]{2012A&A...542A..41C}
{Chauvin}, G., {et~al.} 2012, \aap, 542, A41

\bibitem[{{Chiang} \& {Brown}(1999)}]{1999AJ....118.1411C}
{Chiang}, E.~I., \& {Brown}, M.~E. 1999, \aj, 118, 1411

\bibitem[{{Close} {et~al.}(2012){Close}, {Males}, {Kopon}, {Gasho}, {Follette},
  {Hinz}, {Morzinski}, {Uomoto}, {Hare}, {Riccardi}, {Esposito}, {Puglisi},
  {Pinna}, {Busoni}, {Arcidiacono}, {Xompero}, {Briguglio}, {Quiros-Pacheco},
  \& {Argomedo}}]{2012SPIE.8447E..0XC}
{Close}, L.~M., {et~al.} 2012, Proc. SPIE, 8447

\bibitem[{{Dumusque} {et~al.}(2012){Dumusque}, {Pepe}, {Lovis},
  {S{\'e}gransan}, {Sahlmann}, {Benz}, {Bouchy}, {Mayor}, {Queloz}, {Santos},
  \& {Udry}}]{2012Natur.491..207D}
{Dumusque}, X., {et~al.} 2012, \nat, 491, 207

\bibitem[{{Forveille} {et~al.}(2011){Forveille}, {Bonfils}, {Delfosse},
  {Alonso}, {Udry}, {Bouchy}, {Gillon}, {Lovis}, {Neves}, {Mayor}, {Pepe},
  {Queloz}, {Santos}, {Segransan}, {Almenara}, {Deeg}, \&
  {Rabus}}]{2011arXiv1109.2505F}
{Forveille}, T., {et~al.} 2011, ArXiv e-prints

\bibitem[{{Hardy}(1998)}]{1998aoat.book.....H}
{Hardy}, J.~W. 1998, {Adaptive Optics for Astronomical Telescopes}

\bibitem[{{Hinz} {et~al.}(2012){Hinz}, {Codona}, {Guyon}, {Hoffmann}, {Skemer},
  {Hora}, {Tolls}, {Boss}, {Weinberger}, {Arbo}, {Connors}, {Durney},
  {McMahon}, {Montoya}, \& {Vaitheeswaran}}]{2012SPIE.8446E..1PH}
{Hinz}, P., {et~al.} 2012, in SPIE, Vol. 8446, SPIE

\bibitem[{{Johns} {et~al.}(2012){Johns}, {McCarthy}, {Raybould}, {Bouchez},
  {Farahani}, {Filgueira}, {Jacoby}, {Shectman}, \&
  {Sheehan}}]{2012SPIE.8444E..1HJ}
{Johns}, M., {et~al.} 2012, Proc. SPIE, 8444

\bibitem[{{Kalas} {et~al.}(2008){Kalas}, {Graham}, {Chiang}, {Fitzgerald},
  {Clampin}, {Kite}, {Stapelfeldt}, {Marois}, \& {Krist}}]{2008Sci...322.1345K}
{Kalas}, P., {et~al.} 2008, Science, 322, 1345

\bibitem[{{Kasper} {et~al.}(2010){Kasper}, {Beuzit}, {Verinaud}, {Gratton},
  {Kerber}, {Yaitskova}, {Boccaletti}, {Thatte}, {Schmid}, {Keller}, {Baudoz},
  {Abe}, {Aller-Carpentier}, {Antichi}, {Bonavita}, {Dohlen}, {Fedrigo},
  {Hanenburg}, {Hubin}, {Jager}, {Korkiakoski}, {Martinez}, {Mesa}, {Preis},
  {Rabou}, {Roelfsema}, {Salter}, {Tecza}, \& {Venema}}]{2010SPIE.7735E..81K}
{Kasper}, M., {et~al.} 2010, Proc. SPIE, 7735, 7735E

\bibitem[{{Kasting} {et~al.}(1993){Kasting}, {Whitmire}, \&
  {Reynolds}}]{1993Icar..101..108K}
{Kasting}, J.~F., {Whitmire}, D.~P., \& {Reynolds}, R.~T. 1993, \icarus, 101,
  108

\bibitem[{{Kopparapu} {et~al.}(2013){Kopparapu}, {Ramirez}, {Kasting}, {Eymet},
  {Robinson}, {Mahadevan}, {Terrien}, {Domagal-Goldman}, {Meadows}, \&
  {Deshpande}}]{2013ApJ...765..131K}
{Kopparapu}, R.~K., {et~al.} 2013, \apj, 765, 131

\bibitem[{{Lafreni{\`e}re} {et~al.}(2007){Lafreni{\`e}re}, {Doyon}, {Marois},
  {Nadeau}, {Oppenheimer}, {Roche}, {Rigaut}, {Graham}, {Jayawardhana},
  {Johnstone}, {Kalas}, {Macintosh}, \& {Racine}}]{2007ApJ...670.1367L}
{Lafreni{\`e}re}, D., {et~al.} 2007, \apj, 670, 1367

\bibitem[{{Lagrange} {et~al.}(2010){Lagrange}, {Bonnefoy}, {Chauvin}, {Apai},
  {Ehrenreich}, {Boccaletti}, {Gratadour}, {Rouan}, {Mouillet}, {Lacour}, \&
  {Kasper}}]{2010Sci...329...57L}
{Lagrange}, A.-M., {et~al.} 2010, Science, 329, 57

\bibitem[{{Leconte} {et~al.}(2010){Leconte}, {Soummer}, {Hinkley},
  {Oppenheimer}, {Sivaramakrishnan}, {Brenner}, {Kuhn}, {Lloyd}, {Perrin},
  {Makidon}, {Roberts}, {Graham}, {Simon}, {Brown}, {Zimmerman}, {Chabrier}, \&
  {Baraffe}}]{2010ApJ...716.1551L}
{Leconte}, J., {et~al.} 2010, \apj, 716, 1551

\bibitem[{{Liu} {et~al.}(2010){Liu}, {Wahhaj}, {Biller}, {Nielsen}, {Chun},
  {Close}, {Ftaclas}, {Hartung}, {Hayward}, {Clarke}, {Reid}, {Shkolnik},
  {Tecza}, {Thatte}, {Alencar}, {Artymowicz}, {Boss}, {Burrows}, {de Gouveia
  Dal Pino}, {Gregorio-Hetem}, {Ida}, {Kuchner}, {Lin}, \&
  {Toomey}}]{2010SPIE.7736E..53L}
{Liu}, M.~C., {et~al.} 2010, Proc. SPIE, 7736, 7736E

\bibitem[{{Lowrance} {et~al.}(2005){Lowrance}, {Becklin}, {Schneider},
  {Kirkpatrick}, {Weinberger}, {Zuckerman}, {Dumas}, {Beuzit}, {Plait},
  {Malumuth}, {Heap}, {Terrile}, \& {Hines}}]{2005AJ....130.1845L}
{Lowrance}, P.~J., {et~al.} 2005, \aj, 130, 1845

\bibitem[{{Macintosh} {et~al.}(2012){Macintosh}, {Anthony}, {Atwood},
  {Barriga}, {Bauman}, {Caputa}, {Chilcote}, {Dillon}, {Doyon}, {Dunn},
  {Gavel}, {Galvez}, {Goodsell}, {Graham}, {Hartung}, {Isaacs}, {Kerley},
  {Konopacky}, {Labrie}, {Larkin}, {Maire}, {Marois}, {Millar-Blanchaer},
  {Nunez}, {Oppenheimer}, {Palmer}, {Pazder}, {Perrin}, {Poyneer}, {Quirez},
  {Rantakyro}, {Reshtov}, {Saddlemyer}, {Sadakuni}, {Savransky},
  {Sivaramakrishnan}, {Smith}, {Soummer}, {Thomas}, {Wallace}, {Weiss}, \&
  {Wiktorowicz}}]{2012SPIE.8446E..1UM}
{Macintosh}, B.~A., {et~al.} 2012, Proc. SPIE, 8446

\bibitem[{{Marois} {et~al.}(2006){Marois}, {Lafreni{\`e}re}, {Doyon},
  {Macintosh}, \& {Nadeau}}]{2006ApJ...641..556M}
{Marois}, C., {Lafreni{\`e}re}, D., {Doyon}, R., {Macintosh}, B., \& {Nadeau},
  D. 2006, \apj, 641, 556

\bibitem[{{Marois} {et~al.}(2008{\natexlab{a}}){Marois}, {Lafreni{\`e}re},
  {Macintosh}, \& {Doyon}}]{2008ApJ...673..647M}
{Marois}, C., {Lafreni{\`e}re}, D., {Macintosh}, B., \& {Doyon}, R.
  2008{\natexlab{a}}, \apj, 673, 647

\bibitem[{{Marois} {et~al.}(2008{\natexlab{b}}){Marois}, {Macintosh}, {Barman},
  {Zuckerman}, {Song}, {Patience}, {Lafreni{\`e}re}, \&
  {Doyon}}]{2008Sci...322.1348M}
{Marois}, C., {Macintosh}, B., {Barman}, T., {Zuckerman}, B., {Song}, I.,
  {Patience}, J., {Lafreni{\`e}re}, D., \& {Doyon}, R. 2008{\natexlab{b}},
  Science, 322, 1348

\bibitem[{{Marois} {et~al.}(2010){Marois}, {Zuckerman}, {Konopacky},
  {Macintosh}, \& {Barman}}]{2010Natur.468.1080M}
{Marois}, C., {Zuckerman}, B., {Konopacky}, Q.~M., {Macintosh}, B., \&
  {Barman}, T. 2010, \nat, 468, 1080

\bibitem[{{Mayor} {et~al.}(2009){Mayor}, {Bonfils}, {Forveille}, {Delfosse},
  {Udry}, {Bertaux}, {Beust}, {Bouchy}, {Lovis}, {Pepe}, {Perrier}, {Queloz},
  \& {Santos}}]{2009A&A...507..487M}
{Mayor}, M., {et~al.} 2009, \aap, 507, 487

\bibitem[{{Murray} \& {Correia}(2010)}]{2010exop.book...15M}
{Murray}, C.~D., \& {Correia}, A.~C.~M. 2010, {Keplerian Orbits and Dynamics of
  Exoplanets}, ed. {Seager, S.}, 15--23

\bibitem[{{Parker} \& {Kavelaars}(2010)}]{2010PASP..122..549P}
{Parker}, A.~H., \& {Kavelaars}, J.~J. 2010, \pasp, 122, 549

\bibitem[{{Roelfsema} {et~al.}(2010){Roelfsema}, {Schmid}, {Pragt}, {Gisler},
  {Waters}, {Bazzon}, {Baruffolo}, {Beuzit}, {Boccaletti}, {Charton}, {Cumani},
  {Dohlen}, {Downing}, {Elswijk}, {Feldt}, {Groothuis}, {de Haan}, {Hanenburg},
  {Hubin}, {Joos}, {Kasper}, {Keller}, {Kragt}, {Lizon}, {Mouillet}, {Pavlov},
  {Rigal}, {Rochat}, {Salasnich}, {Steiner}, {Thalmann}, {Venema}, \&
  {Wildi}}]{2010SPIE.7735E.144R}
{Roelfsema}, R., {et~al.} 2010, Proc. SPIE, 7735

\bibitem[{{Skemer} \& {Close}(2011)}]{2011ApJ...730...53S}
{Skemer}, A.~J., \& {Close}, L.~M. 2011, \apj, 730, 53

\bibitem[{{Traub}(2012)}]{2012ApJ...745...20T}
{Traub}, W.~A. 2012, \apj, 745, 20

\bibitem[{{Traub} \& {Oppenheimer}(2011)}]{2011exop.book..111T}
{Traub}, W.~A., \& {Oppenheimer}, B.~R. 2011, {Direct Imaging of Exoplanets},
  ed. S.~{Piper}, 111--156

\bibitem[{{Vigan} {et~al.}(2012){Vigan}, {Patience}, {Marois}, {Bonavita}, {De
  Rosa}, {Macintosh}, {Song}, {Doyon}, {Zuckerman}, {Lafreni{\`e}re}, \&
  {Barman}}]{2012A&A...544A...9V}
{Vigan}, A., {et~al.} 2012, \aap, 544, A9

\bibitem[{{Vogt} {et~al.}(2012){Vogt}, {Butler}, \&
  {Haghighipour}}]{2012arXiv1207.4515V}
{Vogt}, S.~S., {Butler}, R.~P., \& {Haghighipour}, N. 2012, ArXiv e-prints

\bibitem[{{von Braun} {et~al.}(2011){von Braun}, {Boyajian}, {Kane}, {van
  Belle}, {Ciardi}, {L{\'o}pez-Morales}, {McAlister}, {Henry}, {Jao}, {Riedel},
  {Subasavage}, {Schaefer}, {ten Brummelaar}, {Ridgway}, {Sturmann},
  {Sturmann}, {Mazingue}, {Turner}, {Farrington}, {Goldfinger}, \&
  {Boden}}]{2011ApJ...729L..26V}
{von Braun}, K., {et~al.} 2011, \apjl, 729, L26

\bibitem[{{Yamamoto} {et~al.}(2008){Yamamoto}, {Kinoshita}, {Fuse}, {Watanabe},
  \& {Kawabata}}]{2008PASJ...60..285Y}
{Yamamoto}, N., {Kinoshita}, D., {Fuse}, T., {Watanabe}, J.-I., \& {Kawabata},
  K. 2008, \pasj, 60, 285

\end{thebibliography}

\end{document}